\documentclass[traditabstract]{aa} % for a referee version
\usepackage{multirow}
\usepackage{float,lscape}
\usepackage{rotating}
\usepackage{varwidth}
\usepackage{textpos}
\usepackage{rotating} % Rotating table
\usepackage{graphicx, epstopdf}
\usepackage{natbib}
\usepackage{amsmath}
\usepackage{mwe,tikz}
\usepackage[percent]{overpic}

\usepackage[para,online,flushleft]{threeparttable}
\usepackage{txfonts}
\usepackage{hyperref}
\def\src{IGR\,J17329-2731}
\def\inte{{\em INTEGRAL}}
\def\xmm{{\em XMM-Newton}}

\def\swift{{\em Swift}}
\def\rosat{{\em ROSAT}}

\def \inte {{\em INTEGRAL}}
\def \xmm {{\em XMM--Newton}}
\def \nustar {{\em NuSTAR}}

\def \ferg {erg cm$^{-2}$ s$^{-1}$}
\def \hcm {\hbox {\ifmmode $ atom cm$^{-2}\else atom cm$^{-2}$\fi}}

\def \apjl {ApJL}
\def \apjs {ApJS}\def \aap {A\&A}

\def \igr {IGR J17329-2731}

\begin{document}

   \title{IGR\,J17329-2731: The birth of a symbiotic X-ray binary}

   \author{E. Bozzo
    \inst{1}
    \and A. Bahramian
    \inst{2}   
    \and C. Ferrigno
    \inst{1}
    \and A. Sanna 
    \inst{3}    
    \and J. Strader 
    \inst{2}         
    \and F. Lewis 
    \inst{4,5}
    \and D. M. Russell
    \inst{6}
    \and T. di Salvo
    \inst{7}
    \and L. Burderi
    \inst{8}  
    \and  A. Riggio  
    \inst{8}  
    \and A. Papitto 
    \inst{9}
    \and P. Gandhi 
    \inst{10}    
    \and P. Romano
    \inst{11}          
    }
    
   \institute{Department of Astronomy, University of Geneva, Chemin d'Ecogia 16, CH-1290 Versoix, Switzerland; \email{enrico.bozzo@unige.ch}
    \and 
    Department of Physics and Astronomy, Michigan State University, East Lansing, MI, USA
    \and 
    Dipartimento di Fisica, Universit\'a degli Studi di Cagliari, SP Monserrato-Sestu km 0.7, 09042 Monserrato, Italy
    \and 
    Faulkes Telescope Project, School of Physics, and Astronomy, Cardiff University, The Parade, Cardiff CF24 3AA, UK
    \and 
    Astrophysics Research Institute, Liverpool John Moores University, 146 Brownlow Hill, Liverpool L3 5RF, UK
    \and 
    New York University Abu Dhabi, PO Box 129188, Abu Dhabi, United Arab Emirates
    \and 
    Dipartimento di Fisica e Chimica, Universit\'a di Palermo, via Archirafi 36 - 90123 Palermo, Italy
    \and
    Universit\'a degli Studi di Cagliari, Dipartimento di Fisica, SP Monserrato-Sestu, KM 0.7, 09042 Monserrato, Italy
    \and 
    INAF – Osservatorio Astronomico di Roma, via Frascati 33, I-00078, Monteporzio Catone (RM), Italy
    \and 
    Department of Physics and Astronomy, University of Southampton, Highfield, Southampton SO17 1BJ, UK
    \and 
    INAF - Osservatorio Astronomico di Brera, via Emilio Bianchi 46, I-23807 Merate (LC), Italy
}

   \date{}

  \abstract{We report on the results of the multiwavelength campaign carried out after the discovery of the \inte\ transient 
  \src.\ The optical data collected with the SOAR telescope allowed us to identify the donor star in this system as a late M giant at a distance 
  of 2.7$^{+3.4}_{-1.2}$~kpc. The data collected quasi-simultaneously with \xmm\ and \nustar\ showed the presence of a modulation 
  with a period of 6680$\pm$3~s in the X-ray light curves of the source. This unveils that the compact object hosted in this system is a slowly 
  rotating neutron star. The broadband X-ray 
  spectrum showed the presence of a strong absorption ($\gg$10$^{23}$~cm$^{-2}$) and prominent emission lines at 6.4~keV, and 7.1~keV. 
  These features are usually found in wind-fed systems, in which the emission lines result from the fluorescence of the X-rays from the accreting 
  compact object on the surrounding stellar wind. The presence of a strong absorption line around $\sim$21~keV in the \nustar\ spectrum suggests 
  a cyclotron origin, thus allowing us to estimate the neutron star magnetic field as $\sim$2.4$\times$10$^{12}$~G. 
  All evidence thus suggests \src\ is a symbiotic X-ray binary.  
  As no X-ray emission was ever observed from the location of \src\ by \inte\ (or other X-ray facilities) 
  during the past 15~yr in orbit and considering that symbiotic X-ray binaries are known to be variable but persistent X-ray sources, 
  we concluded that \inte\ caught the first detectable X-ray emission from \src\ 
  when the source shined as a symbiotic X-ray binary. The \swift\,XRT monitoring performed up to $\sim$3~months after the discovery 
  of the source, showed that it maintained a relatively stable X-ray flux and spectral properties.}   
  

   \maketitle

\section{Introduction}
\label{sec:intro}

\src\ is an X-ray transient that was discovered by \inte\ on 2017 August 13 \citep{postel17}. At discovery, the source was observed to undergo 
a fast X-ray flare, lasting for about 2~ks and reaching a flux of $\sim$60~mCrab in the 20-40~keV energy range. 
No previously known X-ray sources were detected within the source location accuracy provided by \inte,\ and follow-up 
observations were carried out with several facilities to determine the nature of \igr.\  

The best position of the source in X-rays was provided thanks to the \swift/XRT observations at 
RA(J$2000)$ = $17^{\rm h} 32^{\rm m} 50\fs28$, Dec(J$2000)=-27^{\circ} 30^{\prime} 04\farcs9$ 
with an associated uncertainty of 3\farcs1 at 90\,\% confidence level \citep{bozzo17}. This was later refined thanks to the discovery of the likely 
optical counterpart of the source, identified through observations performed by the 2 m Faulkes Telescope North 
(Maui, Hawaii) and South (Siding Spring, Australia) and the Las Cumbres Observatory (LCO) 
1 m telescopes at Cerro Tololo, Chile. A variable source was revealed within the XRT error circle that had brightened  
by more than 4 magnitudes in the R band since 1991 and was similarly bright in 1976 \citep{russell17}. Observations acquired with the 
SOAR/GOODMAN spectrograph suggested that the optical star is a cool M giant \citep[SyXB;][]{arash17}. 
This suggested that \src\ is a symbiotic X-ray binary \citep[SyXB; see, e.g.,][]{walter2015}, i.e., a relatively rare system in which a 
compact object (likely a neutron star; NS) accretes from the slow wind of its evolved companion. 

We report on all X-ray data collected after the discovery of the source and on the available 
SOAR/GOODMAN observations in order to confirm the nature of \src.\ We also report on a non-detection of the 
source in an archival \rosat/PSPC observation, which is the only other X-ray observation (aside from the \inte\ pointings) 
performed toward the direction of \src\ before its discovery. Throughout this paper, we provide all uncertainties 
on measured parameters at 90~\% confidence level, unless stated differently. We used version 12.9.1p of the {\sc Xspec} 
software \citep{xspec} for all X-ray spectral analyses.

\section{X-ray data}

\subsection{\inte\ data}
\label{sec:integral}

The field around \src\ was observed with \inte\ from the discovery up to 2017 October 27, covering the satellite revolutions 
1850-1878. After t fhis period, the source was no longer observable by \inte\ owing to Sun constraints. 
We analyzed all the publicly available \inte\ data using version 10.2 of the Off-line Scientific Analysis software   
(OSA) distributed by the International Space Development Conferences (ISDC) \citep{courvoisier03}.
The \inte\ observations are divided into science windows (SCWs), 
i.e., pointings with typical durations of $\sim$2-3~ks. We only included SCWs in which the source was located
to within an off-axis angle of 4~deg from the center of the JEM-X field of view (FoV) in the JEM-X\ analysis \citep{lund03}. 
For IBIS/ISGRI \citep{ubertini03,lebrun03}, we retained all SCWs for which the source 
was located within an off-axis angle of 12~deg from the center of the instrument FoV. These choices allowed us to 
minimize the instruments calibration uncertainties\footnote{http://www.isdc.unige.ch/integral/analysis}.  

\begin{figure}
  \includegraphics[width=9.0cm]{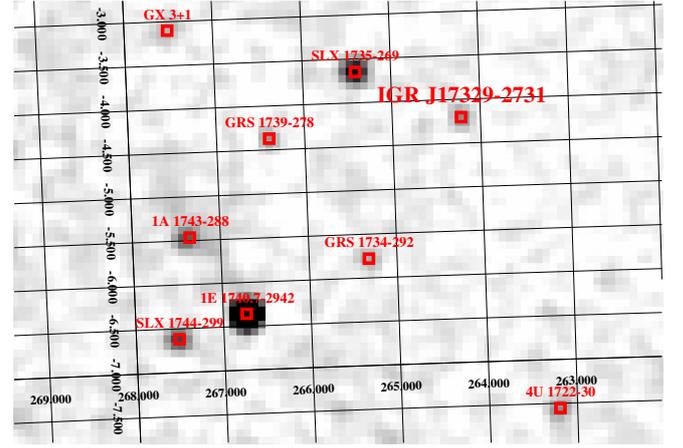}
  \caption{Zoom of the IBIS/ISGRI mosaic of the region around \src,\ as obtained from the combined 
  SCWs in revolutions 1850-1859 (excluding the SCW 35 in revolution 1850). The source is detected at 
  a significance of $10\sigma$.}   
  \label{fig:mosa}
\end{figure}

\begin{figure}
  \includegraphics[width=6.2cm, angle=-90]{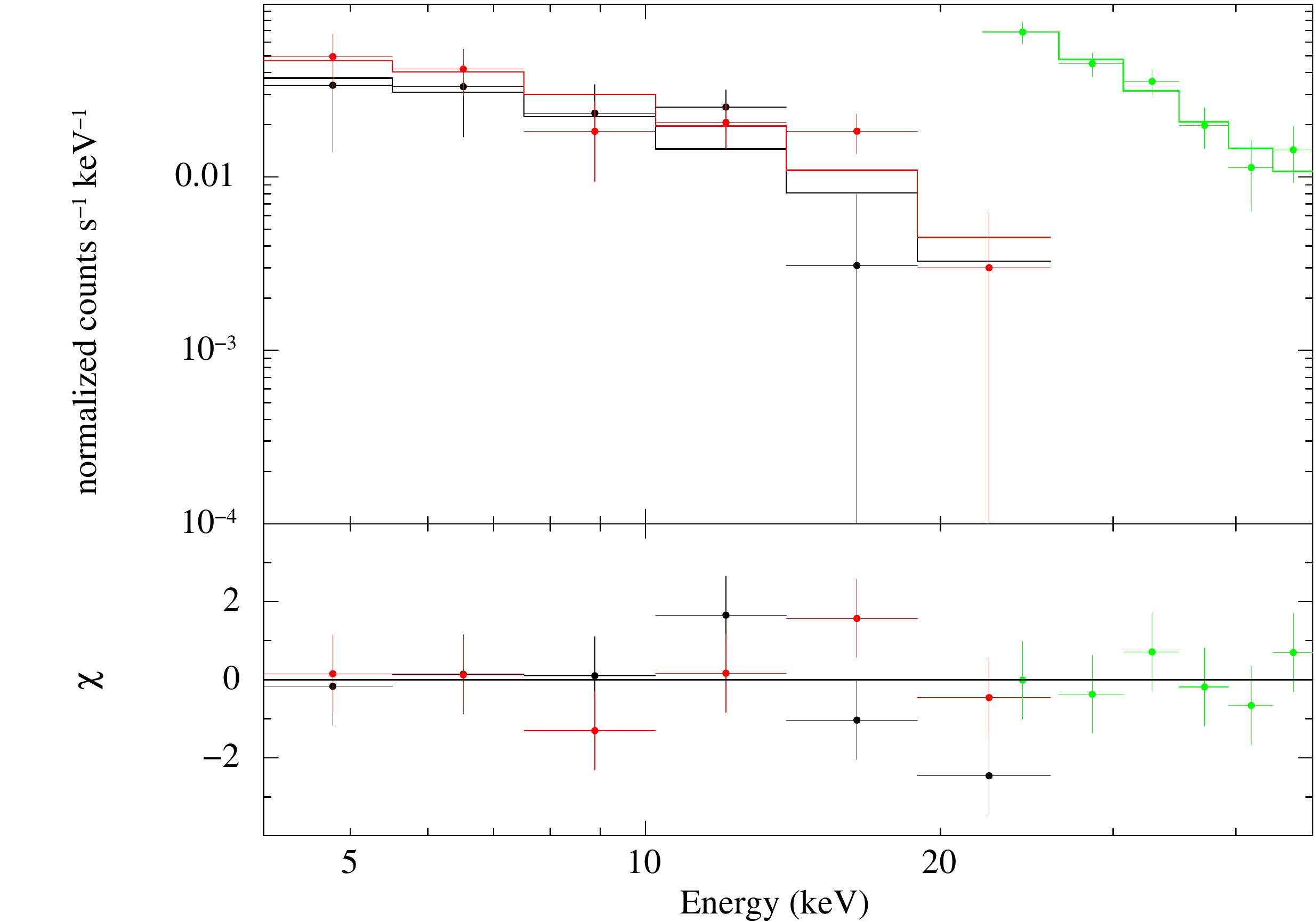}
  \caption{JEM-X1 (black), JEM-X2 (red), and ISGRI (green) spectra of \src\ extracted by using all combined data across 
  revolutions 1850-1878 (excluding SCW 35 in revolution 1850). The best fit model is obtained with an absorbed power law (see text for details). 
  The residuals from the fit are shown in the bottom panel.}   
  \label{fig:int_spe}
\end{figure}

We extracted first the IBIS/ISGRI and JEM-X mosaics dividing the entire 
observational period in two parts: from revolution 1850 to 1859 \citep[excluding the SCW 35 in revolution 1850, where the source was observed to 
undergo a bright flare;][]{postel17}, and from revolution 1871 to 1878. 
\src\ was detected in the IBIS/ISGRI 20-100~keV mosaic of the first period (effective exposure 163.5~ks) at a significance of 
$10\sigma$ and an average count rate of 1.3$\pm$0.1~cts~s$^{-1}$. 
We show the zoom of this mosaic around the position of \src\ in Fig.~\ref{fig:mosa}. 
In the second mosaic (effective exposure 200.0~ks), the detection significance 
was $7.8\sigma$ and the average count rate 0.9$\pm$0.1~cts~s$^{-1}$. 
For the JEM-X instruments, we measured a detection significance 
of $3\sigma$ and an average count rate of 0.6$\pm$0.2~cts~s$^{-1}$ in the first mosaic (3-35~keV; effective exposure 22.8~ks). 
For the second mosaic (effective exposure 25.0~ks), we obtained in the same energy range a detection significance of $2\sigma$ and an average count rate of 
0.4$\pm$0.2~cts~s$^{-1}$. Given the relatively low detection significance of the source in the \inte\ data, 
no meaningful timing analysis was possible and we  
extracted a single spectrum integrating over the entire exposure time available (excluding SCW 35 in revolution 1850)  for ISGRI and the two JEM-X instruments. 
These spectra (Fig.~\ref{fig:int_spe}) could be well fit ($\chi_{\rm red}^2$/d.o.f.=1.20/13) with a simple power-law model (tbabs*pow in {\sc Xspec}). 
We fixed in the fit the absorption column density to the value measured by \xmm\ (see Sect.~\ref{sec:spectra}) and obtained a power-law photon index of 
2.4$\pm$0.4. We included in the fit normalization constants between the two JEM-X and ISGRI, which turned out to be compatible with unity. 
The measured 3-20~keV and 20-100~keV X-ray fluxes were 8.8$\times$10$^{-11}$~\ferg and 7.1$\times$10$^{-11}$~\ferg, respectively. 
For completeness, we also tested that the same model used to described the average \swift/XRT spectrum in Sect.~\ref{sec:swift} would give compatible 
results once applied to the \inte\ data (to within the large uncertainties due to the low statistics of these data). 
The JEM-X1 and JEM-X2 light curves with a time resolution of 2~s were extracted from all available SCWs in revolutions 1850-1878 
to search for type-I X-ray bursts, but no significant detections were found.  
\begin{figure}
  \includegraphics[width=6.1cm, angle=-90]{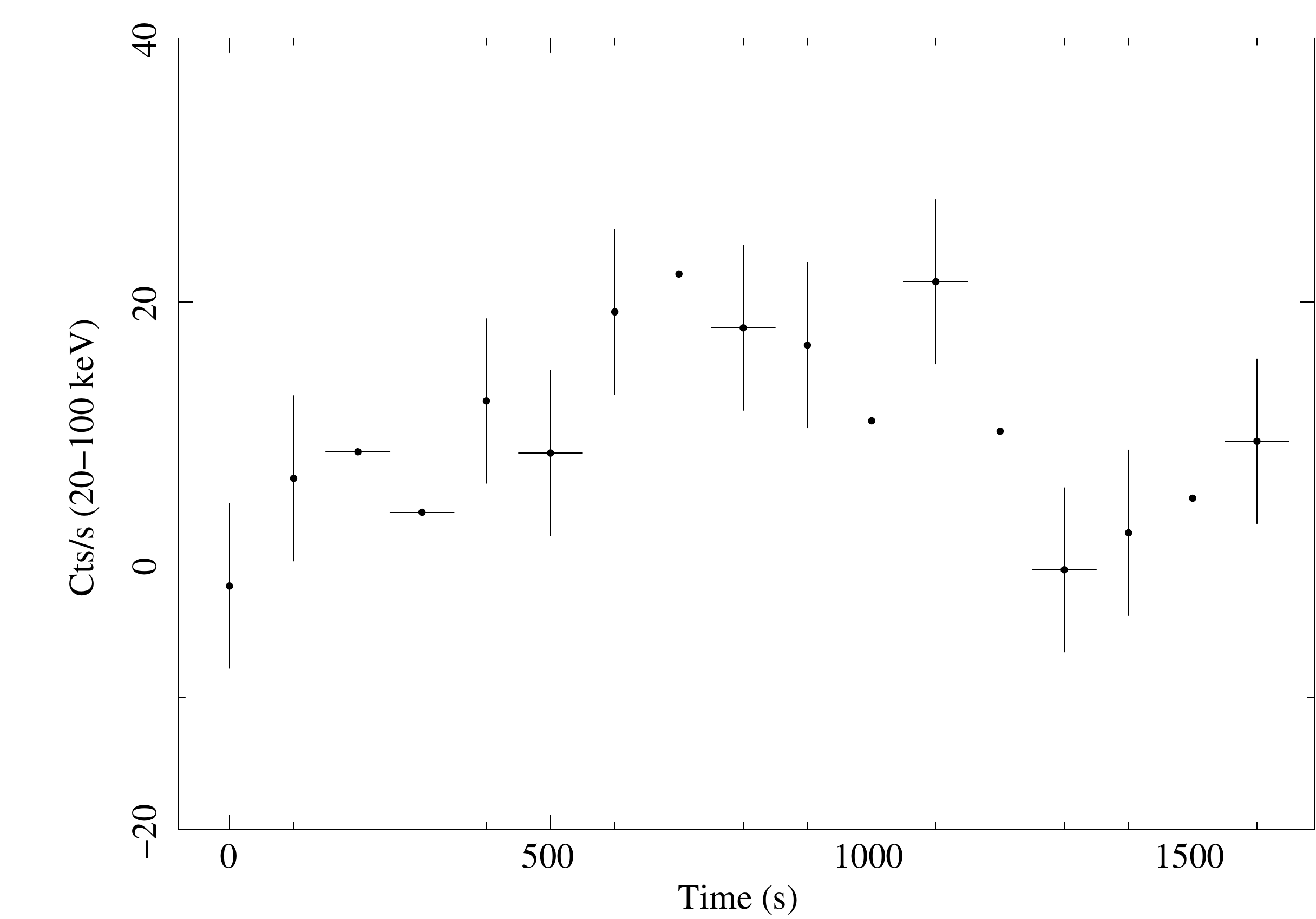}
  \caption{ISGRI light curve of \src\ during SCW 35 (20-100~keV) showing the flare displayed by the source. 
  The time resolution is 100~s.}   
  \label{fig:isgrilc}
\end{figure}

We analyzed separately the \inte\ data obtained from SCW 35 in revolution 1850. 
As reported by \citet{postel17}, the source was located at the very rim of the JEM-X FoV during the observational time covered by this SCW 
(2017 August 13 from 16:35 to 17:05 UTC) and not detected by this instrument. In the corresponding IBIS/ISGRI data, 
the average source count rate was of 7.3$\pm$1.0~cts~s$^{-1}$ and the light curve extracted  with 100~s resolution 
(Fig.~\ref{fig:isgrilc}) suggests that the source underwent a flare. 
The ISGRI spectrum extracted from this SCW could be well fit with a simple power law with photon index 2.9$_{-1.3}^{+1.6}$, 
thus indicating no significant changes in the source spectral properties compared to the longer integration spectrum described before 
(to within the relatively large uncertainties).   

Finally, we exploited the entire archive of the INTEGRAL data to search for possible historical detections of the source by IBIS/ISGRI. 
We collected a total of 15172 SCWs within which \src\ was located less than 12~deg away from the instrument aim point. 
We inspected the IBIS/ISGRI images of all SCWs and found no previous significant detections of the source. 
From the IBIS/ISGRI mosaic built using all these data (effective exposure 31.3~Ms), we estimated a $3\sigma$ upper limit on any hard X-ray emission 
from the source before its discovery of 1.3$\times$10$^{-12}$~\ferg (20-100~keV; i.e., $\sim$0.1~mCrab).

\subsection{\xmm\ data}
\label{sec:xmm}

The \xmm\ observation of \src\ began on 2017-08-30 04:48 UT and lasted until 15:55 UTC (OBSID: 0795711701), 
providing a total exposure time of 38.7~ks for the two EPIC-MOS and 36.7~ks for the EPIC-pn. 
The pn was operated in timing mode, while the MOS1 was operated in full frame and the MOS2 in small window mode. 
Data were also collected with the two grating instruments RGS1 and RGS2, but their data were not usable given the large extinction in 
the direction of the source (see below). No flaring background intervals were recorded, and thus the entire exposure time 
could be used for the scientific analysis. 

\begin{figure}
 \centering
 \includegraphics[width=5.8cm,angle=-90]{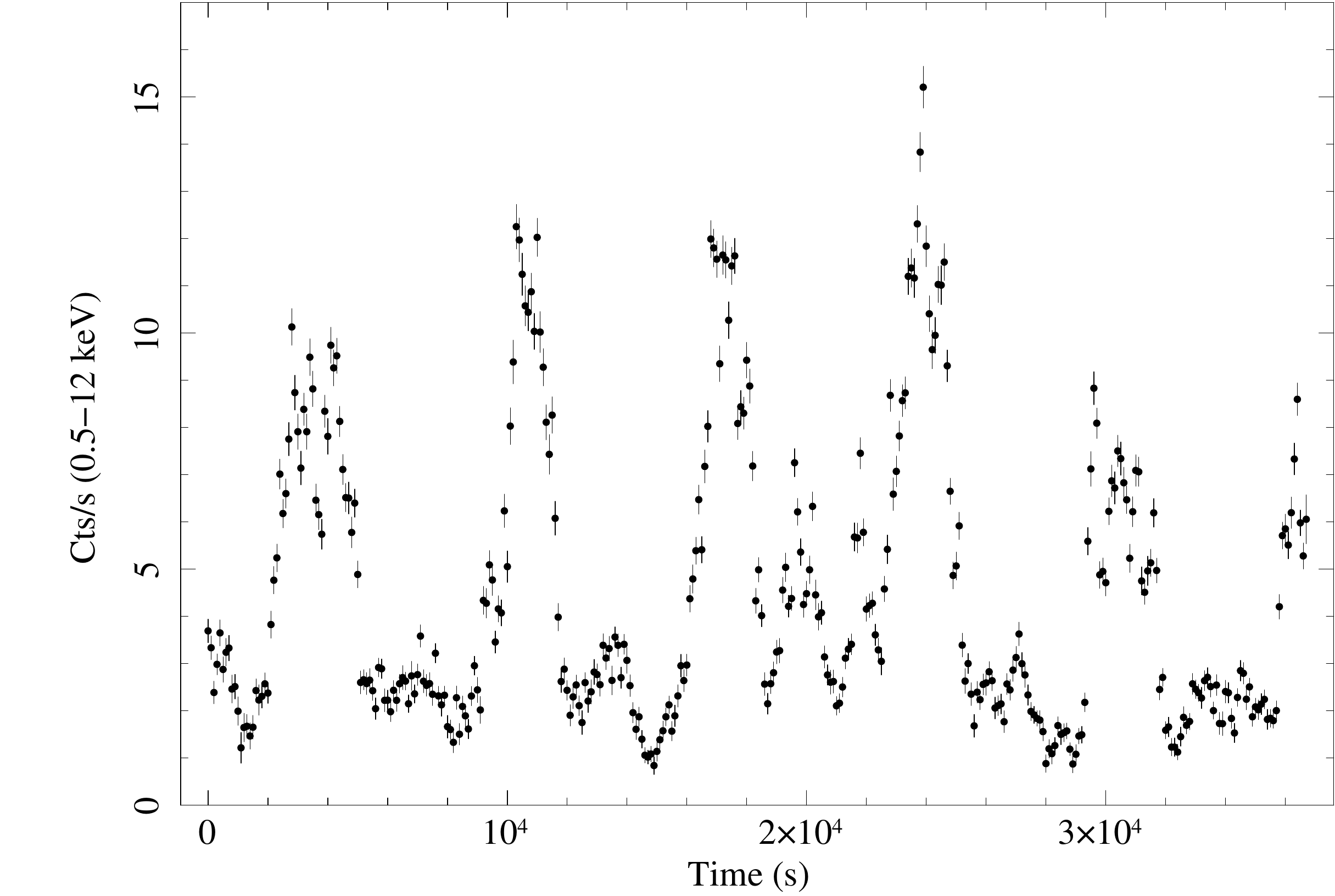}
 \includegraphics[width=6cm,angle=-90]{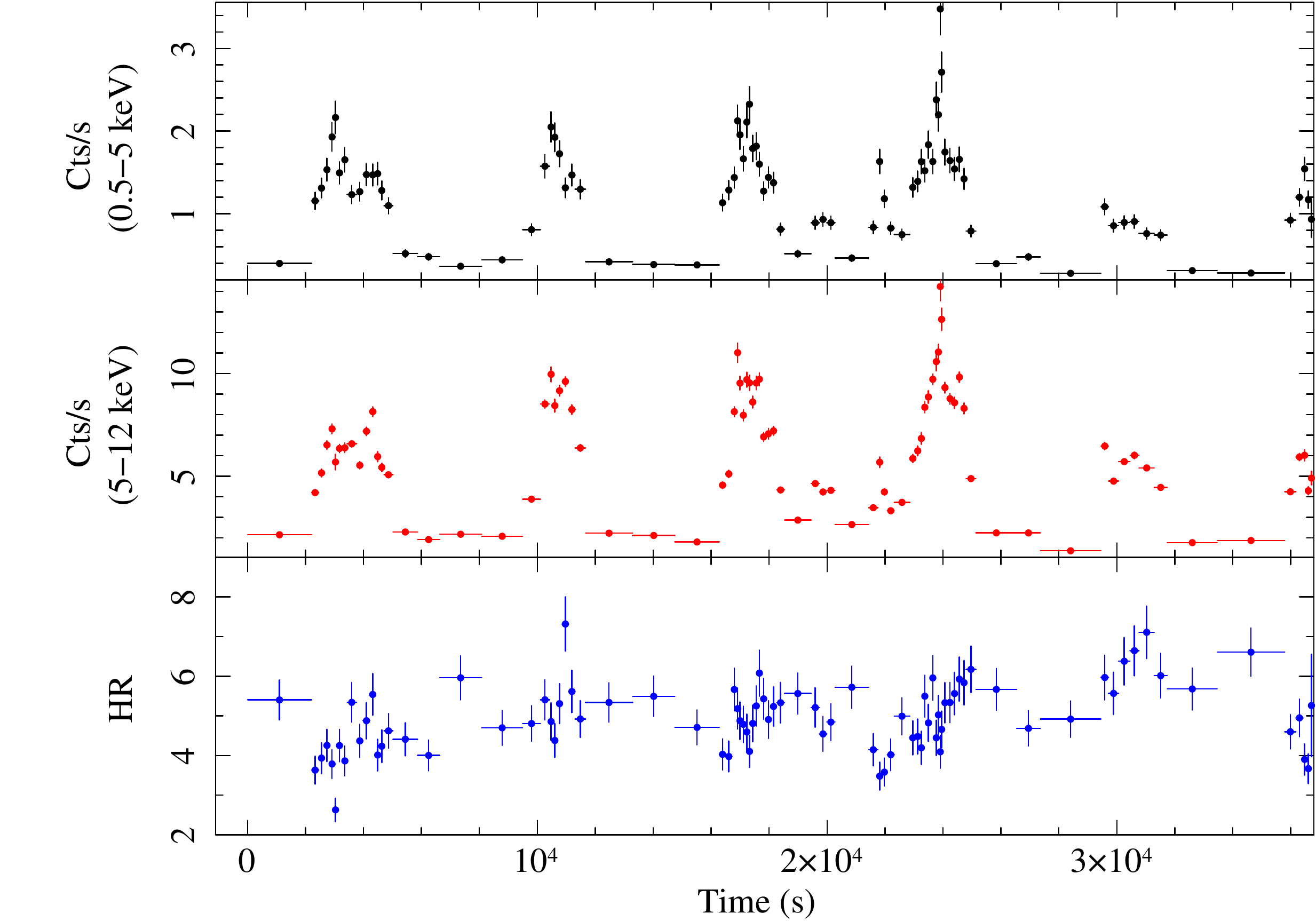}
  \caption{{\em Top}: 0.5-12~keV pn light curve of \src\ with a time resolution of 100~s. {\em Bottom}: The energy-resolved pn light curve 
  of \src\ (top and middle panels) and the correspondingly computed HR (bottom panel).}
  \label{fig:xmm_lc}
\end{figure}

All observation data files (ODFs) were processed by using the \xmm\ Science Analysis System 
(SAS 16.1) following standard procedures\footnote{http://www.cosmos.esa.int/web/xmm-newton/sas-threads}. 
The regions used for the extraction of the source spectra and light curves were chosen 
for all instruments to be centered on the best known position of \src\ (see Sect.~\ref{sec:intro}). 
The background spectra and light curve extraction regions were chosen to lie in a portion of the instrument 
FoV free from the contamination of the source emission. 
We show in the top plot of Fig.~\ref{fig:xmm_lc} the background-corrected light curve of the source in the 0.5-12~keV 
energy range. The source displays a clearly regular modulation with a period of $\sim$6680~s, which we interpret as the pulse period 
of the compact object in \src\ and investigate in more detail in Sect.~\ref{sec:timing}. The same modulation is 
observed in the MOS1 and MOS2 light curves. The average and peak count rate of the source along this modulation was not sufficiently 
high to cause any pileup in the pn and MOS2 data. However, MOS1 data were significantly affected by pileup (verified with the   
SAS {\sc epaplot} task) and thus discarded for further analysis as they could not provide any significant improvement of the results. 

In Fig.~\ref{fig:xmm_lc}, we also show the energy resolved pn light curves of the source and the corresponding hardness ratio (HR). 
This was computed with an adaptive rebinning \citep[see, e.g.,][]{bozzo13b}, achieving in each soft 
time bin a signal-to-noise ratio (S/N) of $\gtrsim$10. No remarkable changes of the HR are recorded that could have indicated 
large spectral variations during the pulse phase. We thus did not perform any pulse phase or HR-resolved spectral analysis. 

We also extracted the pn and MOS2 spectra using the entire exposure time available. The pn and MOS2 spectra were fit together with the quasi-simultaneous 
\nustar\ data and the results are reported in Sect.~\ref{sec:spectra}.

\subsection{\nustar\ data}
\label{sec:nustar}

\src\ was observed by \nustar\ from 2017 August 29 at 15:35 to August 30 at 02:36 (UTC). The Target of Opportunity observation 
(OBSID~90301012002) was triggered as close as possible to the \xmm\ observation (see Sect.~\ref{sec:xmm}). 
After having applied to the \nustar\ data all the good time intervals (GTI) accounting for the Earth occultation and 
the South Atlantic Anomaly passages, we obtained an effective exposure time of 20.8~ks for both the  
focal plane modules A and B (FPMA and FPMB). The data were processed via {\sc nustardas v1.5.1} and the latest 
calibration files available (v.20171002). The source spectra and light curves were extracted from a 80~arcsec circle centered on the source, 
while the background products were extracted from a region with a similar extension but centered on a region free from the contamination 
of the source emission. Various extraction regions were also used for the source and background products to verify that 
none of the timing and spectral features could be affected by some specific choices (see Sect.~\ref{sec:timing} and \ref{sec:spectra}). 

\begin{figure}
 \centering
 \includegraphics[width=5.5cm,angle=-90]{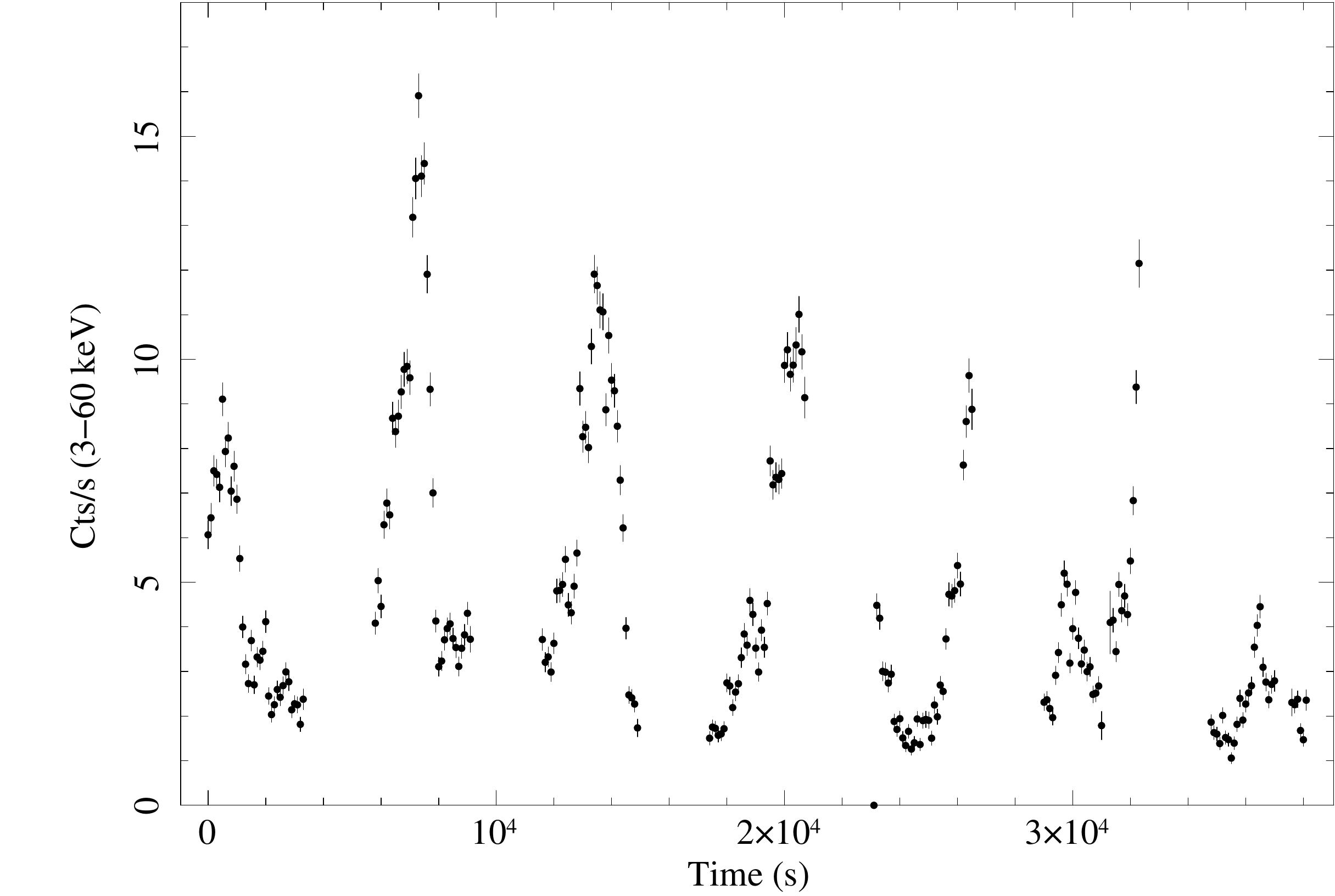}
  \caption{\nustar\ FPMA light curve of \src\ in the 3-60~keV energy range.}
  \label{fig:nustar_lc}
\end{figure}

The FPMA and FPMB light curves of the source displayed a clear modulation with a period of $\sim$6680~s (Fig.~\ref{fig:nustar_lc}), 
similar to that observed from the \xmm\ data (see Sect.~\ref{sec:xmm}). 
The timing analysis of the \nustar\ data is presented in Sect.~\ref{sec:timing}. The FPMA and FPMB source spectra were 
rebinned in order to have at least 100 photons per energy bin and fit together with the \xmm\ spectra in Sect.~\ref{sec:spectra}.  

\subsection{\swift\ data}
\label{sec:swift}

\src\ was observed by the Neil Gehrels \swift\ Observatory \citep{burrows05} from three days after the discovery up to 2017 October 26, 
when the source entered into a few months-long Sun constraint.  
The monitoring program requested from \swift\ provide deeper coverage during the first few weeks 
after the discovery (up to several 1~ks-long observations per week) and was relaxed in the following period 
as the source showed a relatively stable flux level and spectral properties. 

\begin{figure}
\centering
 \includegraphics[width=4.0cm,angle=-90]{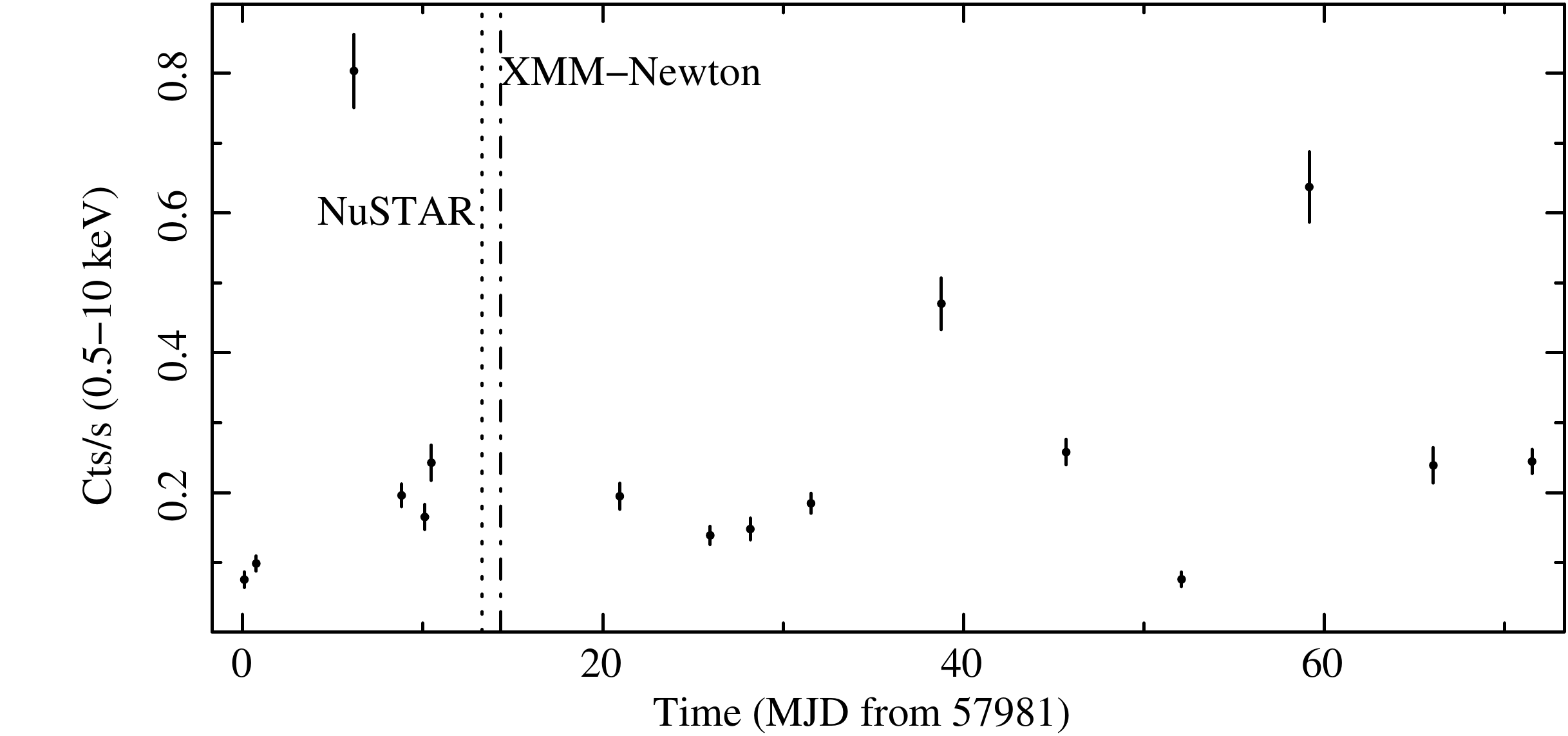}
  \caption{Light curve obtained from the \swift\,/XRT monitoring campaign performed from 2017 August 16 (57981~MJD) up to 
  2017 October 26 (58052~MJD). Each point of the light curve corresponds to the average count rate of the source during every 
  XRT observation. The vertical lines indicate the mid-time of the \xmm\ (dot-dashed line) and \nustar\ (dotted line) observations.} 
  \label{fig:xrt_lc}
\end{figure}

The XRT data were analyzed via the standard software ({\sc Heasoft} v6.22.1) and the latest  
calibration files available (CALDB 20170501).  
All data were processed and filtered with {\sc xrtpipeline} (v0.13.4). 
The source displayed an average count rate of 0.155$\pm$0.006~cts~s$^{-1}$ (0.5-10~keV)  throughout the
campaign and no data were significantly affected by pileup. The source events were extracted from a circular region 
with a radius of 20 pixels (where 1 pix corresponds to $\sim2\farcs36$), while 
background events were extracted from a source-free region with a similar radius.   
We show in Fig.~\ref{fig:xrt_lc} the long-term background subtracted XRT light curve in the 0.5-10~keV energy band 
corrected for point spread function losses and vignetting. The XRT spectra extracted from each observation  
could be well fit with an absorbed power-law model. This more complex spectral model did not improve the fits and resulted in 
poorly constrained spectral parameters. A log of all XRT observations used with the corresponding spectral fit 
results is reported in Table~\ref{tab:xrt}. The marginally noticeable correlation between the measured values of the absorption 
column density and the power-law photon index is due to the combination of the limited statistics of the single XRT pointings and 
the narrow bandpass of the instrument.  
\begin{table*} 
\centering 
 \caption{Log of all {\em Swift}/XRT observations used in the present paper. We also report for each observation 
 the results obtained from the spectral analysis. In all cases, the best fit model is an absorbed power law ({\sc tbabs*pow} 
 in {\sc Xspec)}, where we used {\em wilm} abundances \citep{wilms00} and {\em vern} cross sections \citep{vern96}. 
 The reported flux is in the 0.5-10~keV energy range and in units of 10$^{-11}$~\ferg (it is not corrected for absorption). We note that a few adjacent observations 
 were merged together to obtain a reasonable statistics to perform the spectral fit. All spectra were fit with the C-statistics \citep{cash79} due to the 
 relatively low number of counts.}      
 \label{tab:xrt} 
 \begin{tabular}{@{}lllllllll@{}} 
 \hline 
 \hline 
 \noalign{\smallskip} 
 Sequence   & Obs.  & Start time  & End time & Exposure & $N_{\rm H}$ & $\Gamma$ & Flux$_{\rm 0.5-10~keV}$ & cstat/d.o.f. \\ 
                  &   Mode   & (UTC)  & (UTC)  & (s) & (10$^{23}$~cm$^{-2}$) &  &  &  \\
  \noalign{\smallskip} 
 \hline 
 \noalign{\smallskip} 
00010244001     &       PC      &       2017-08-16 02:26:50      &  2017-08-16 02:45:55 &      \multirow{2}{*}{1973}   &       \multirow{2}{*}{4.9$\pm$4.0}    &       \multirow{2}{*}{-0.9$^{+1.4}_{-1.6}$}  &      \multirow{2}{*}{2.9$^{+0.8}_{-0.6}$}    & \multirow{2}{*}{39.6/23} \\
00010244002     &       PC      &       2017-08-16 17:03:06      &  2017-08-16 20:02:21        &               &                                &                               &                                 & \\
00010244003     &       PC      &       2017-08-22 03:47:02      &  2017-08-22 05:30:03        &       960     &       5.1$^{+1.7}_{-1.5}$     &       1.6$^{+1.1}_{-1.0}$         &       16.0$\pm$3.0            & 55.8/49 \\
00010244004     &       PC      &       2017-08-24 16:43:09      &  2017-08-24 23:11:00        &       872     &       6.6$^{+3.2}_{-2.7}$     &       1.4$^{+1.7}_{-1.5}$         &       4.9$^{+1.4}_{-0.9}$     & 28.8/26 \\
00010244005     &       PC      &       2017-08-25 23:00:00      &  2017-08-26 06:39:00        &       601     &       2.5$^{+2.8}_{-2.1}$     &       -0.9$^{+1.5}_{-1.4}$         &       5.1$^{+1.7}_{-1.3}$     & 10.7/14 \\
00010244006     &       PC      &       2017-08-26 08:06:00      &  2017-08-26 14:59:00        &       862     &       5.3$^{+4.7}_{-4.4}$     &       -0.4$^{+1.7}_{-1.6}$         &       7.8$^{+2.5}_{-2.0}$     & 13.7/15 \\
00010244007     &       PC      &       2017-09-05 21:29:00      &  2017-09-05 23:17:00        &       659     &       4.2$^{+2.8}_{-2.4}$     &       1.8$^{+1.8}_{-1.6}$         &       3.6$^{+1.2}_{-0.8}$     & 25.9/18 \\
00010244008     &       PC      &       2017-09-10 22:35:00      &  2017-09-10 22:53:00        &       952     &       6.6$^{+3.6}_{-3.1}$     &       1.1$^{+1.9}_{-1.7}$         &       3.6$^{+1.3}_{-0.9}$     & 18.0/19 \\
00010244009     &       PC      &       2017-09-13 01:36:00      &  2017-09-13 06:28:00        &       717     &       8.2$^{+4.8}_{-3.8}$     &       3.2$^{+2.8}_{-2.3}$         &       2.7$^{+1.2}_{-0.7}$     & 19.7/14 \\
00010244010     &       PC      &       2017-09-16 12:43:00      &  2017-09-16 13:02:00        &       1100    &       3.3$^{+1.4}_{-1.6}$     &       1.5$^{+1.2}_{-1.1}$         &       3.4$^{+0.8}_{-0.6}$     & 51.8/29 \\
00010244011     &       PC      &       2017-09-23 15:10:00      &  2017-09-23 21:48:00        &       897     &       2.6$^{+2.6}_{-2.5}$     &       0.7$^{+1.3}_{-1.2}$         &       8.7$^{+1.9}_{-1.7}$     & 23.3/32 \\
00010244012     &       PC      &       2017-09-30 16:25:00      &  2017-09-30 16:42:00        &       955     &       6.3$^{+2.7}_{-2.1}$     &       2.7$^{+1.7}_{-1.5}$         &       5.1$^{+1.5}_{-1.0}$     & 32.2/36\\
00010244013     &       PC      &       2017-10-07 01:12:00      &  2017-10-07 02:57:00        &       \multirow{2}{*}{1508}   &       \multirow{2}{*}{1.6$^{+0.8}_{-0.7}$}    &       \multirow{2}{*}{0.7$\pm$0.8}  &      \multirow{2}{*}{5.3$^{+1.1}_{-0.9}$}    & \multirow{2}{*}{72.8/44} \\
00010244014     &       PC      &       2017-10-14 03:51:00      &  2017-10-14 04:03:00        &               &               &               &               & \\
00010244015     &       PC      &       2017-10-21 00:26:00      &  2017-10-21 01:56:00        &       \multirow{2}{*}{1381}   &       \multirow{2}{*}{4.5$^{+1.5}_{-1.3}$}    &       \multirow{2}{*}{1.5$^{+1.0}_{-0.9}$}  &      \multirow{2}{*}{5.1$\pm$0.8}    & \multirow{2}{*}{53.7/57} \\
00010244016     &       PC      &       2017-10-26 12:28:00      &  2017-10-26 12:45:00        &               &               &               &               & \\
  \noalign{\smallskip}
  \hline
  \end{tabular} 
  \end{table*}

The XRT data showed that, following the \inte\ discovery, the source remained at a relatively stable X-ray flux 
and did not show significant spectral variability. The limited flux and 
spectral variability measured across the various observations are typical of wind-fed systems and in the case of \src\ are also 
related to the modulation of the X-ray emission by the long spin period of the source (see Sect.~\ref{sec:timing}).   
We also fit with an absorbed power-law model the source spectrum 
extracted by stacking together all XRT data (effective exposure time 13.2~ks). The fit gave a poor result ($\chi^2_{\rm red}$/d.o.f.=1.7/78), and thus we improved 
the description of these data ($\chi^2_{\rm red}$/d.o.f.=1.1/76) adding a high energy cutoff ({\sc highecut} in {\sc Xspec}; 
see also Sect.~\ref{sec:spectra}). We measured in this case an absorption column density of (2.3$\pm$0.8)$\times$10$^{23}$~cm$^{-2}$, a 
power-law photon index of $\Gamma$=-0.5$\pm$0.8, a folding energy $E_{\rm fold}$=3.3$^{+3.4}_{-1.2}$~keV, a cutoff energy of 
$E_{\rm cut}$=6.4$\pm$0.3~keV, and an average 0.5-10~keV flux of (4.5$\pm$0.3)$\times$10$^{-11}$~\ferg. We noticed that fixing the 
value of the folding energy to that measured from the combined \xmm\ and \nustar\ data ($E_{\rm fold}$=14.5~keV)  
would not significantly affect the quality of the spectral fit ($\chi^2_{\rm red}$/d.o.f.=1.1/77). Furthermore, the other parameters would be 
in good agreement with those reported in Sect.~\ref{sec:spectra}. In particular, we obtained in this case 
$\Gamma$=0.5$\pm$0.3, $N_{\rm H}$=(3.3$\pm$0.5)$\times$10$^{23}$~cm$^{-2}$, and $E_{\rm cut}$=6.4$\pm$1.4~keV. This further confirmed that the source did not 
undergo significant spectral variations during the entire observational period monitored with \swift,\ \xmm,\ and \nustar.\

\subsection{\rosat\ data}
\label{sec:rosat}

A \rosat/PSPC \citep{rosat} observation was carried out in the direction of \src\ on 1992 February 29 for a total exposure time of 3.7~ks. 
The processed image in the 0.1-2.4~keV energy range was downloaded from the {\em HEASARCH} archive, together with the corresponding exposure map. 
\src\ was not detected in this observation and we determined a $3\sigma$ upper limit on the source count rate of 0.003~Cts~s$^{-1}$ via the 
tool\footnote{https://heasarc.gsfc.nasa.gov/docs/rosat/faqs/data\_src\_faq3.html} {\sc sosta} available within {\sc ximage} 
({\sc heasoft} v.6.22.1). We converted this count rate into a flux using the online {\sc webpimms} tool and assuming a spectral model comprising an absorption 
column density of 5$\times$10$^{23}$~cm$^{-2}$ and a power law with a photon index ranging between $\Gamma$=0.5 and $\Gamma$=3.0 (see Table~\ref{tab:xrt}). 
Because of the limited energy coverage of the instrument and the strong absorption local to the source, we obtained unconstraining   
$3\sigma$ upper limits on the 0.5-10~keV flux within the range (0.4-7)$\times$10$^{-4}$~erg~cm$^{-2}$~s$^{-1}$.

\subsection{\xmm\ and \nustar\ timing analysis}
\label{sec:timing}

Given the evident modulation of the source light curve in the \xmm\ data, we first corrected the arrival times of the photons recorded by the pn camera 
to the solar system barycenter using the best available position of the source and then carried out a detailed timing analysis. 
We created a Lehay-normalized power density spectrum (PDS) with Nyqvist frequency of 1024~Hz by averaging $\sim$16ks data intervals. 
The PDS shows a prominent feature at $\sim$6680~s and a second peak at twice this value. We interpret these features 
as the pulse period of the source and its harmonic. To further investigate this periodicity, 
we performed epoch-folding search of the whole \xmm\ observation using 80 phase bins and exploring the frequency space around the value 
obtained from the analysis of the PDS with steps of 10$^{-7}$~Hz for a total of 1001 steps. The best X-ray pulse profile is obtained using   
a period of 6692$\pm$4~s (1~$\sigma$ c.l.). To estimate the uncertainty on the period we performed Monte Carlo simulations 
by generating 100 datasets with equal exposure, count rate, and pulse profile as the observed 
data\footnote{We verified {\em a posteriori} that using a larger number of realizations ($\gg$100) did not significantly affect the 
results.}. For each of these datasets, we then applied 
the same procedure described above to obtain the best period value of the periodicity. Finally, we estimated the period uncertainty as the 
standard deviation of the best period values distribution. 

\begin{figure}
 \centering
 \includegraphics[width=9.1cm]{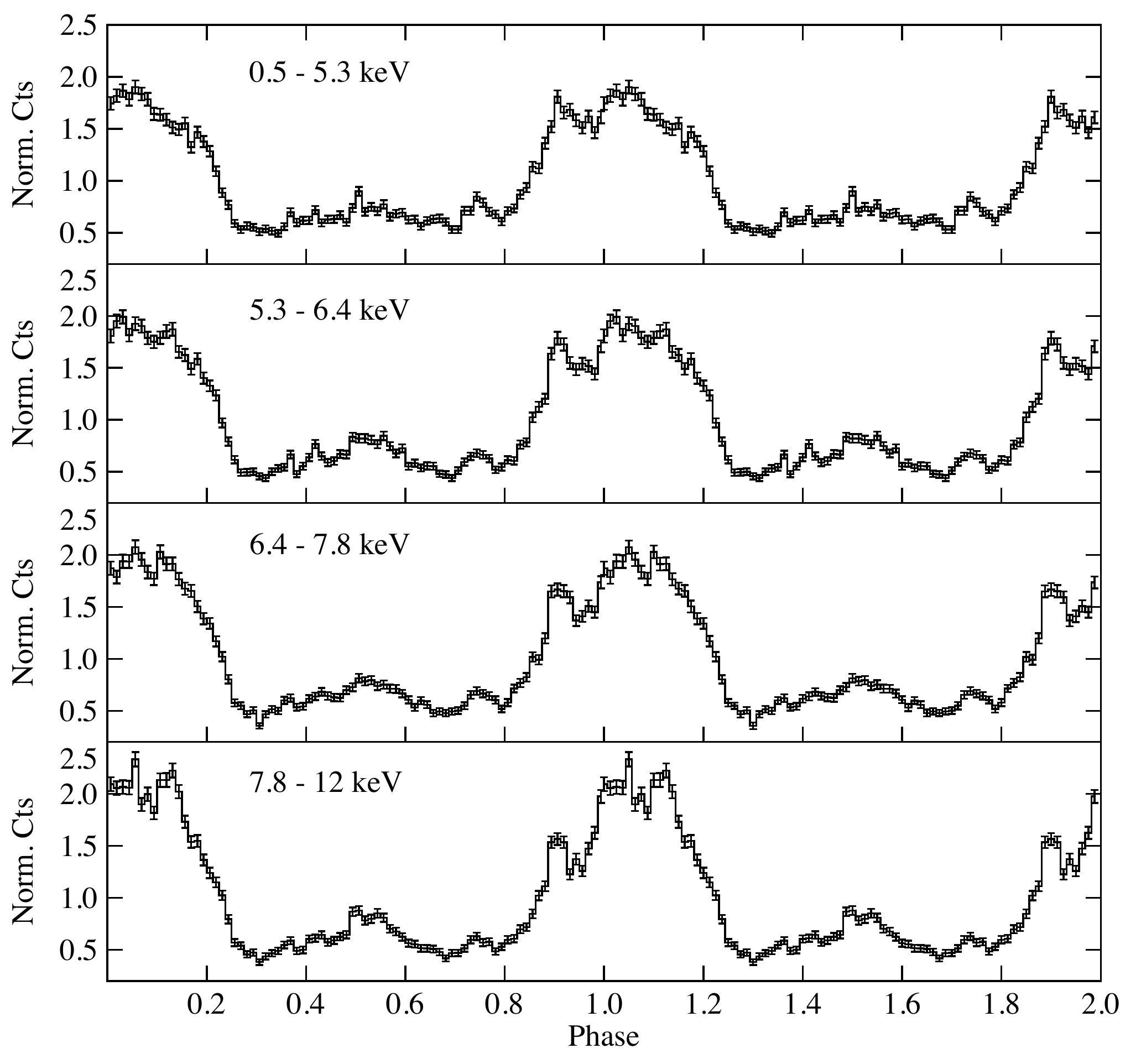}
  \caption{Pulse profiles of \src\ as determined from the EPIC-pn data. The best determined pulse period  
  used to obtain the profiles is 6692$\pm$4~s (1~$\sigma$ c.l.).}
  \label{fig:pulse}
\end{figure}

\begin{figure}
 \centering
 \includegraphics[width=9.1cm]{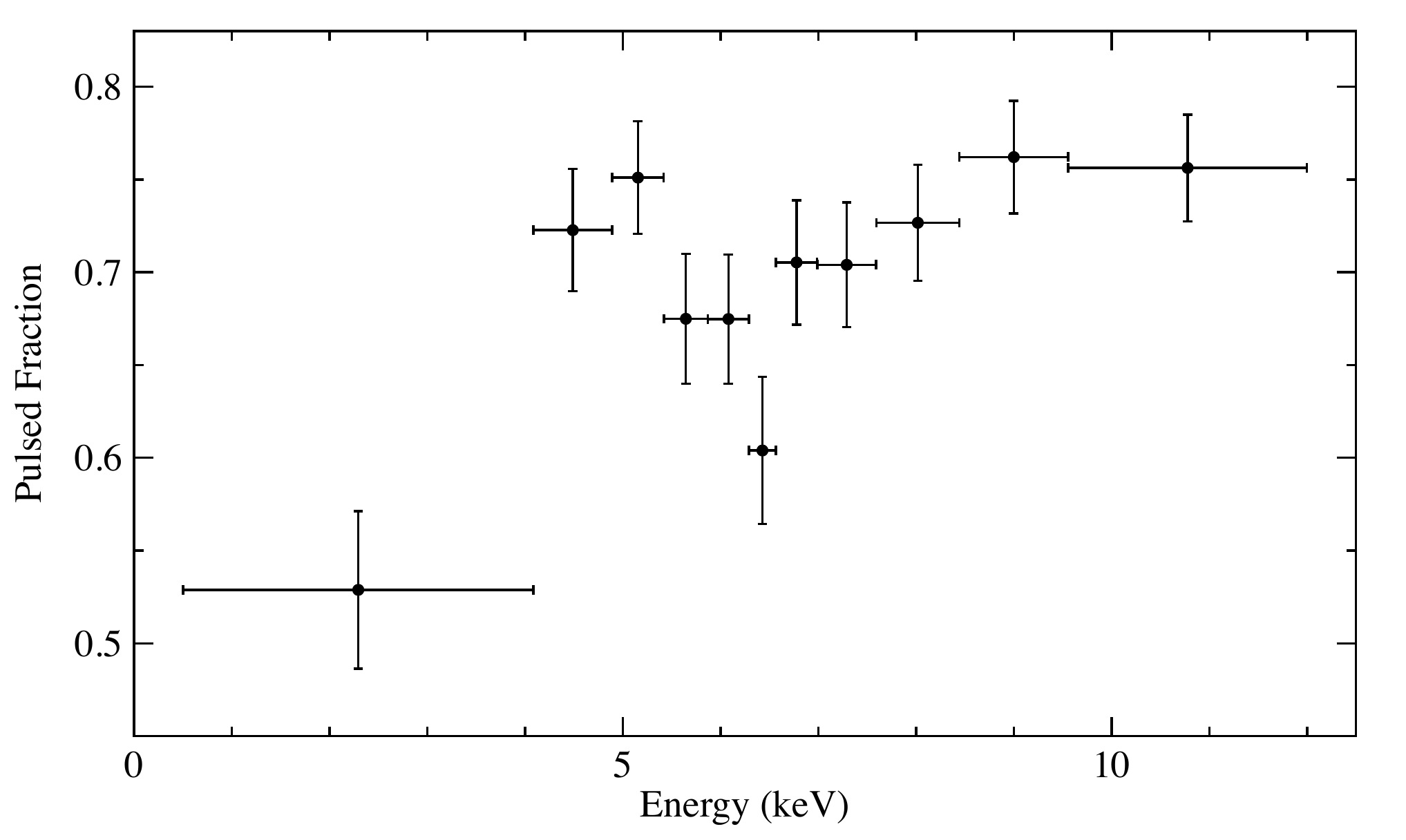}
  \caption{Pulsed fraction as a function of the energy derived from the EPIC-pn data.}
  \label{fig:fract}
\end{figure}

We also investigated the average pulse profile as a function of energy by dividing 
the \xmm\ energy range in four intervals characterized by roughly the same number of photons. Figure~\ref{fig:pulse} shows the average pulse profiles for the 
selected energy bands. No significant differences are observed within the energy range covered by the pn. Finally, we estimated the pulse 
fraction $f$ as a function of energy selecting 11 statistically equivalent intervals (see Fig.~\ref{fig:fract}) with the equation 
\begin{equation}
f=\frac{max(P)-min(P)}{max(P)+min(P)}
,\end{equation}
where $P$ is the pulse profile. The choice of statistically equivalent intervals allowed us to 
ensure that the measured changes of $f$ as a function of energy are an intrinsic property of the source. 
Our findings suggest that the pulse fraction increases as a function of energy, and there is an indication of a possible 
decrease around the fluorescence iron line energy ($\sim$6.4~keV, see below). This is similar to what is observed in 
other X-ray pulsars and particularly in wind-fed systems (see also Sect.~\ref{sec:discussion}). 

As the \nustar\ data are affected by regular interruptions due to the occultation of the source by the Earth, we used a different technique to 
determine the best source period.  The source FPMA and FPMB light curves with 100~s resolution (3-60~keV) were first normalized to the corresponding average count rates, 
combined together, and then modeled with a Bayesian approach to derive the posterior distribution of the fitting parameters. 
We assumed as a baseline model a constant plus four Fourier harmonics, such that the total number of 
parameters are the constant $D$, the relative amplitudes and phases of the harmonics ($A_i$ and $\phi_i$), and the spin period $P$. 
To prevent boundary issues, amplitudes were varied from zero to one and phases from zero to 1.5. 
The intrinsic source variability was accounted for in the method by multiplying  
the amplitudes of the Fourier harmonics by a factor sampled from a normal distribution centered at unity (represented as $\xi$ below). 
In addition, we introduced an overall intrinsic scatter in the source emission by sampling its count rate from a normal distribution centered 
on the expected value. The amplitudes of these distributions are indicated as $\sigma_p$ and $\sigma$ and are parameters of the 
model with a non-constraining prior. To summarize, the model can be expressed as $\mathcal{N}(c, \sigma)$, where 
\begin{equation}
c= D \left[ 1+ 2 \xi \sum_{i=1}^4 A_i \cos \left( 2\pi \left( i \frac{t}{P} - \phi_i \right) \right) \right]  
\end{equation}
and 
\begin{equation}
\xi = \mathcal{N}(1, \sigma_p). 
\end{equation}
In the equation above, $\mathcal{N}(m, s)$ represents a normal distribution with mean $m$ and standard deviation $s$. We used 
a Gibbs sampler from \emph{JAGS}\footnote{\url{https://sourceforge.net/projects/mcmc-jags/files/}} 
and three chains for consistency. Following a burn-in phase of 2000 realizations, we sampled 1000 elements 
from each chain with a length of 10000 realizations. We inspected the parameter distribution and found no significant 
degeneracies between them. The best spin period obtained with this method is 6696$\pm$15\,s at $1\sigma$ c.l., and we measured 
a variability of the total source emission of $\sigma$=59$\pm$4\% (all uncertainties here are given at $1\sigma$ c.l.). 
The parameter $\sigma_p$ could not be constrained in the fit and was fixed to 0, verifying that this did not significantly affect the results. 
The same technique applied to the \xmm\ data provided a period that is compatible with that reported above 
(6694$\pm$13~s). In the case of the \xmm\ data we measured from the fit $\sigma_p$=39$\pm$4\% and $\sigma$=86$\pm$7\%. 

\begin{figure}
 \centering
 \includegraphics[width=9.1cm]{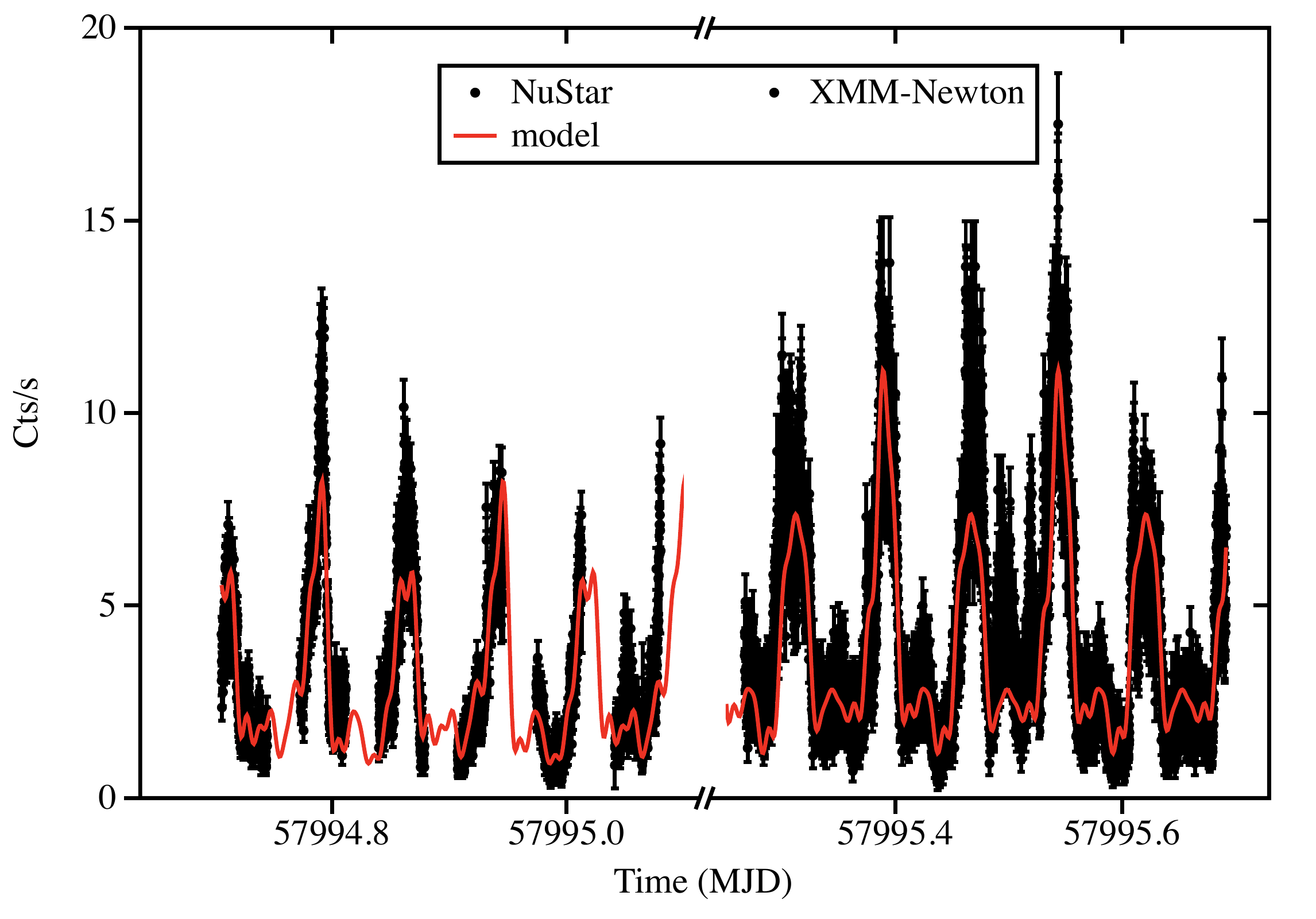}
  \caption{Results of the fit to the \xmm\ (right side of the timeline) and \nustar\ (left side of the timeline) light curves 
  of \igr\ with a superposition of sinusoidal functions with harmonically related 
  periods. The red curve represents the best fit model and the most accurately derived spin period of the source from the combined 
  \xmm\ + \nustar\ dataset is $6680\pm3$~s (1~$\sigma$).}
  \label{fig:curvefit}
\end{figure}

We also adopted for further confirmation a third technique, fitting both the \xmm\ and \nustar\ light curves  
with a superposition of sinusoidal function with periods harmonically related $f(x)$=$a$+$\sum_{n=1}^{13}$$b_n$$\sin{\left(\frac{2\pi(t-\phi_n)}{nP}\right)}$, 
where $a$, $b_n$, and $\phi_n$ represent a constant, the amplitude, and the time phase of the sinusoidal functions. We obtained 
a spin period of $6688\pm4$~s (1~$\sigma$), $6675\pm5$~s (1~$\sigma$), and $6680\pm3$~s (1~$\sigma$) for \xmm,\ \nustar,\ and the two datasets combined; 
in the combined fit, we only forced the period to be the same among the two datasets but left all other parameters free to vary (Fig.~\ref{fig:curvefit}). 
These values are well in agreement with the periods determined through the various techniques above and we thus consider in the following 
the last value as our best and most accurate estimate of the pulse period from \igr.\   

\begin{figure}
 \centering
 \includegraphics[width=3.4cm,angle=-90]{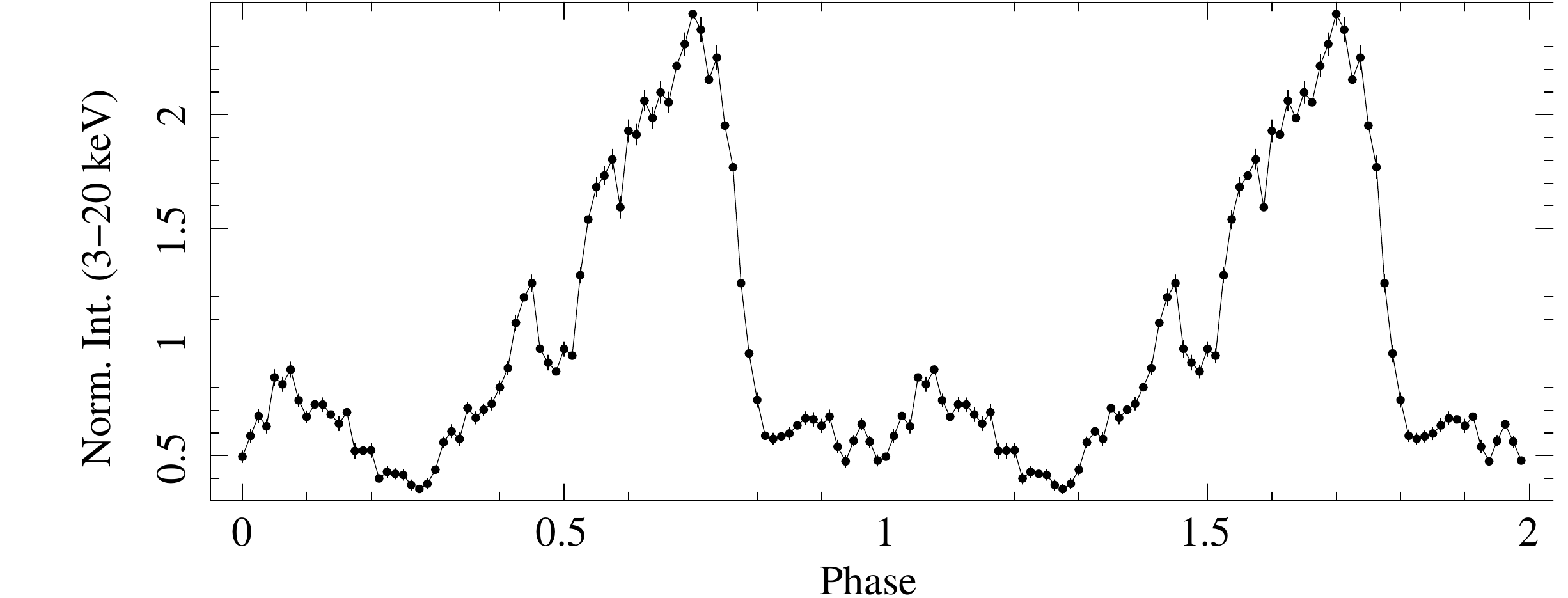}
 \includegraphics[width=3.4cm,angle=-90]{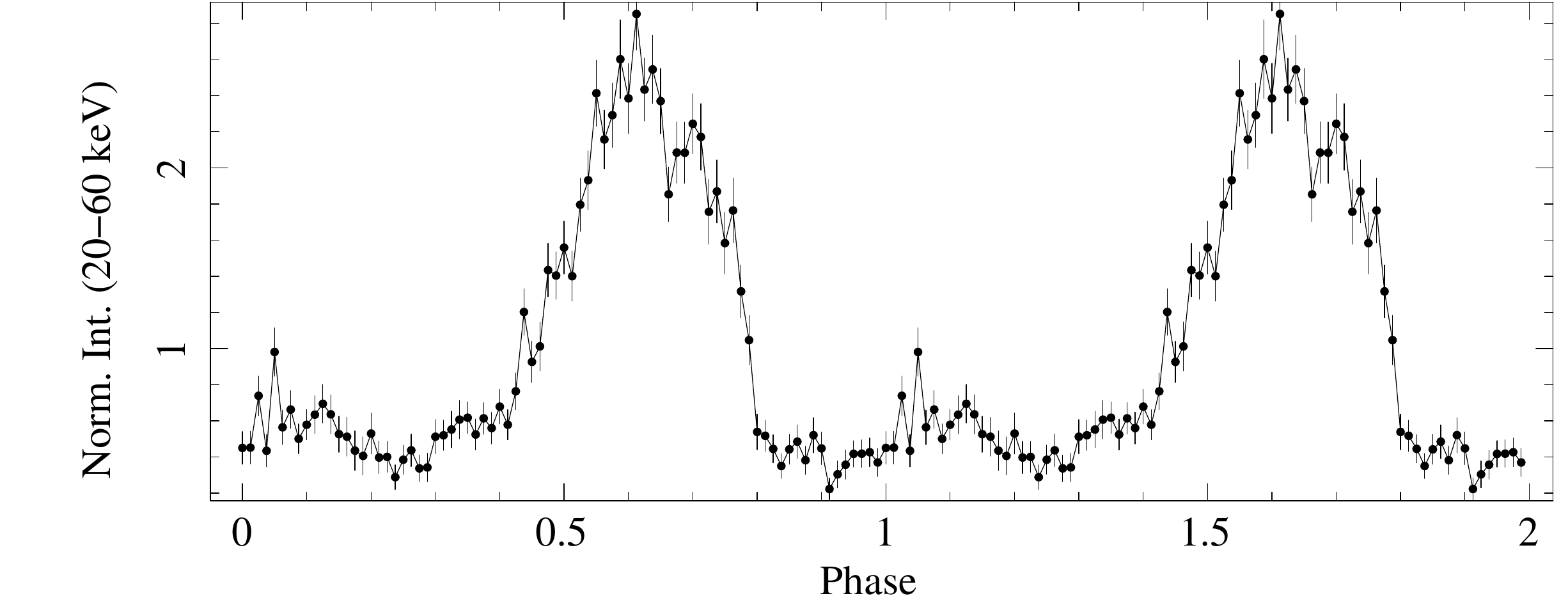}
  \caption{Pulsed profiles obtained from the \nustar\ FPMA data in various energy ranges (3-20~keV and 20-60~keV). 
  The profiles were obtained from the source light curves folded on the best determined pulse period of 6680$\pm$3\,s (1~$\sigma$).}
  \label{fig:nustar_pulse}
\end{figure}

We show in Fig.~\ref{fig:nustar_pulse} the source energy-resolved folded pulse profiles obtained from the \nustar\ 
data, using the most accurate period determined so far of 6680$\pm$3~s. We only show FPMA data, but equivalent results were obtained from the FPMB; 
the phase 0 was set to 57994~MJD. There seem to be no remarkable differences between the lower (3-20~keV) 
and higher energy (20-60~keV) pulse profiles. We also could not find noticeable differences by changing the energy bands 
of these profiles slightly. 

\subsection{\xmm\ and \nustar\ combined spectral analysis}
\label{sec:spectra}

\begin{figure*}
\centering
 \includegraphics[width=14cm,angle=-90]{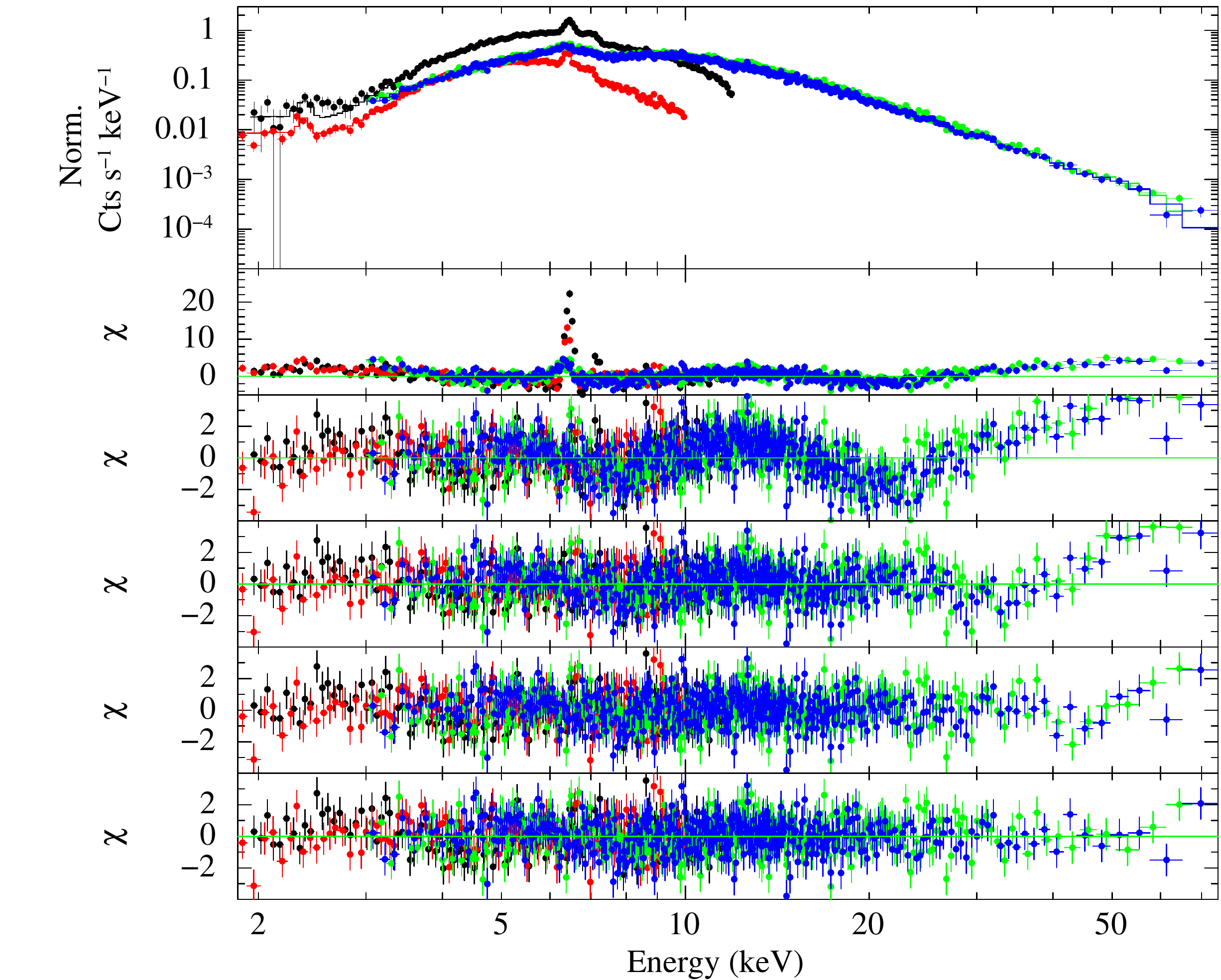}
  \caption{Quasi-simultaneous \xmm\ and \nustar\ spectra of \src.\ The EPIC-pn is in black, the EPIC-MOS2 in red, the FPMA in green, and the FPMB in 
  blue. The best fit to the data is obtained with the model {\sc const*tbabs*tbpcf*gabs*gabs*gabs* (gaussian+gaussian+gaussian+highecut*powerlaw)}. 
  The residuals from the best fit are shown in the bottom panel. The second panel from the top shows the residuals obtained using a simple absorbed 
  power law with a high energy cutoff ({\sc tbabs*highecut*powerlaw}).
  The third panel from the top shows the residuals obtained using a simple absorbed power law with a 
  high energy cutoff and the three lines at 2.3~keV, 6.4~eV, and 7.1~keV ({\sc tbabs*(gaussian+gaussian+gaussian+highecut*powerlaw)}). 
  The fourth panel from the top shows the residuals obtained by including a {\sc gabs} component with a centroid energy of $\sim$20~keV, while the fifth  
  panel reports the residuals obtained when a second  {\sc gabs} component with a centroid energy of $\sim$30~keV is accounted for in the model as well.} 
  \label{fig:spectra}
\end{figure*}

In this section, we report on the detailed broadband spectral analysis of the X-ray emission from \src,\ exploiting the 
quasi-simultaneous \xmm\ and \nustar\ data. These spectra could not be described by a simple absorbed power-law model ($\chi^{2}_{\rm red}$/d.o.f.=7.89/1051), 
where large residuals were left at both the lower and higher ends of the covered energy range, and around 6-7~keV (see Fig.~\ref{fig:spectra}). 

We improved the fits using a number of different phenomenological models. The best description of the source continuum emission was achieved 
with an absorbed power-law including a cutoff at the higher energies and a partial absorber ({const*tbabs*tbpcf*highecut*pow} in {Xspec}, 
where $const$ is a normalization constant included to take into account inter-calibration uncertainties between the different instruments and the 
fact that the data were not strictly simultaneous). We used {\em wilm} abundances and 
{\em vern} cross sections. The value of the first absorption column density was fixed to 3$\times$10$^{21}$~cm$^{-2}$, 
which agrees with the expected Galactic value in the direction of the source\footnote{https://heasarc.gsfc.nasa.gov/cgi-bin/Tools/w3nh/w3nh.pl?}.  
We included three Gaussians in the fit to take into account the presence of evident emission lines at 2.3~keV, 
6.4~keV and 7.1~keV, together with a partial covering model to take into account the residuals below 2~keV (spectral model 1 in Table~\ref{tab:spectra}). 
The three emission lines correspond to the S K$\alpha$, Fe K$\alpha$, and Fe K$\beta$, respectively \citep[see, e.g.,][and references therein]{fuerst11}. 

As can be seen from Fig.~\ref{fig:spectra}, this model has not yet provided an acceptable result 
($\chi^{2}_{\rm red}$/d.o.f=1.74/1037). In particular, residuals resembling the shape of absorption features were still left 
around $\sim$20~keV, $\sim$30~keV, and $\sim$40~keV. We verified that these residuals could not be decreased with alternative source and 
background extraction regions, 
leading to the conclusion that they are intrinsic to the source. To the best of our knowledge, there are no instrumental features 
at the energies of the identified absorption lines. We thus included a first absorption feature ({\sc gabs} in {Xspec}) at 20~keV, 
significantly improving the fit up to $\chi^{2}_{\rm red}$/d.o.f.=1.25/1035. A second {\sc gabs} component at 30~keV further improved the fit 
up to $\chi^{2}_{\rm red}$/d.o.f.=1.16/1033. A third {\sc gabs} component with a centroid energy around $\sim$40~keV led only to a marginal 
improvement of the fit ($\chi^{2}_{\rm red}$/d.o.f.=1.14/1031). In the final fit we fixed the widths of the three {\sc gabs} components 
to the best fit values obtained by the previous fits to avoid continuum modeling with the absorption features 
\citep[this is a common practice for similar fits of complex X-ray pulsar spectra; see, e.g.,][]{ferrigno09, muller13}.  
We verified that leaving these parameters free to vary in the fit did not significantly affect the results. 
A nearly equivalent fit could be obtained by changing the partial covering component with a low temperature blackbody ({\sc diskbb} in {\sc Xspec}). 
As the low energy turnover of the \xmm\ and \nustar\ spectra was characterized by an evidently different curvature, we left free to vary in the fit 
either the partial covering fraction or the normalization of the disk blackbody component between the two datasets. 
The normalization of the 6.4~keV iron line was also significantly 
different between \xmm\ and \nustar,\ and thus left free to vary in the fit. 
We report all the results of the spectral fits in Table~\ref{tab:spectra}.  
In all cases, the normalization constants turned out to be compatible with unity and are thus not reported explicitly in the table. 

For completeness, we also performed a fit to the broadband spectrum using the {\sc cyclabs} component in {\sc Xspec} instead of the 
three {\sc gabs} components. This model provided an equivalently good fit to the data, but the width of the fundamental line (now centered at  
19.7$^{+0.3}_{-0.7}$~keV) and the first harmonic turned out to be rather large (8.9$^{+0.3}_{-0.8}$~keV and 12.7$^{+3.3}_{-4.7}$~keV, 
respectively). On the one hand, it is interesting to note that only two lines in harmonic ratio are required with this alternative model 
(no residuals are left around 30~keV). On the other hand, the two broad features seem to be used by the {\sc Xspec} fit routine to cover  
a large portion of the source continuum emission and thus we did not consider the model with a {\sc cyclabs} component as a convincingly 
viable alternative. 

\begin{figure}
\centering
 \includegraphics[width=7.8cm,angle=-90]{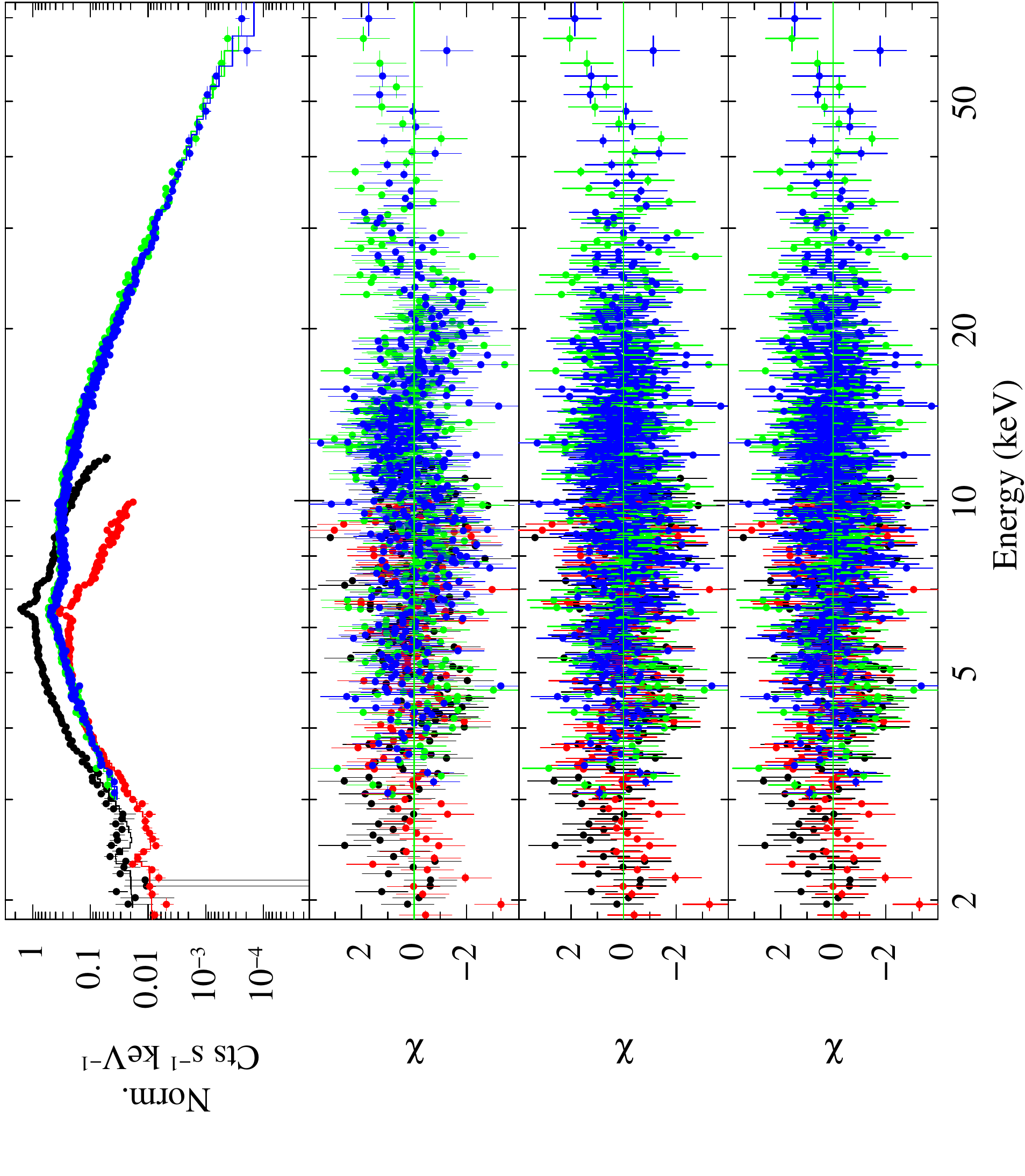}
  \caption{Same as Fig.~\ref{fig:spectra} but in the case of the best fit model obtained with a combination of a partial covering, a hot {\sc bbodyrad}, and a 
  {\sc comptt} component (plus the three Gaussians at 6.4, 7.1, and 2.3~keV). The residuals in the second panel from the top correspond to the case in which no {\sc gabs} 
  is included in the fit, while those in the third panel from the top show the improvement obtained by including in the fit an absorption feature around 21~keV. The bottom 
  panel shows the residuals from the best fit, where a possible second absorption component around 40~keV wa included in the model as well.} 
  \label{fig:spectra2}
\end{figure}

A number of other models were tested as well, including a combination of a partial covering with a blackbody 
({\sc bbodyrad} in {\sc Xspec}) and a Comptonized plasma model ({\sc comptt} in {\sc Xspec}). In the first case, indicated as 
``Model 2'' in Table~\ref{tab:spectra}, the blackbody radius turns out to be unphysically too low for this component to 
be associated with the emission from an accretion disk around the compact object in \src\ (see Sect.~\ref{sec:discussion}). 
For the model using a {\sc comptt} component, indicated as ``Model 3'' in Table~\ref{tab:spectra}, we linked in the fit the temperature of the 
{\sc bbodyrad} component to be the same between the \xmm\ and \nustar\ data, while the normalization of this component was left free 
to vary among the two datasets. As the {\sc comptt} covers mainly the high energy part of the spectrum, all component parameters 
were linked between the \xmm\ and \nustar\ data. In this model, the {\sc bbodyrad} component has a hot temperature ($\sim$1.4~keV) 
and a small radius ($\lesssim$1~km), suggesting a hot spot origin on the NS surface. The seed photon temperature needed in order for the 
{\sc comptt} component to cover properly a large fraction of the source hard continuum emission is much larger than that of the {\sc bbodyrad} 
component, and linking the two temperatures did not result in an acceptable fit (featuring ``S'' shaped residuals). The {\sc comptt} component 
is usually adopted to describe the Comptonized emission in the accretion columns of X-ray pulsars, and it cannot be excluded that the seed photons 
have a different origin with respect to the NS hot spot or could be part of these photons already up-scattered to higher energies. The drawback 
of this assumption is that the resulting plasma temperature of the model is roughly an order of magnitude higher ($kT$$\sim$120~keV) than what is usually 
observed in other X-ray pulsars and especially in symbiotic X-ray binaries observed with instruments providing a wide X-ray energy coverage 
\citep[see, e.g.,][]{masetti07b,enoto14,kita}. 
The interesting outcome of the fit with this alternative model is that 
only the {\sc gabs} component at $\sim$21~keV remains highly significant, while residuals at $\sim$40~keV are barely noticeable.   
The addition of a {\sc gabs} component at $\sim$40~keV only led to a marginal improvement of the fit from $\chi^2_{\rm red}$/d.o.f.=1.16/1031 
to $\chi^2_{\rm red}$/d.o.f.=1.15/1029, but still permitted us to get rid of the residuals above $\sim$40~keV (see Fig.~\ref{fig:spectra2}). 
No residuals were left around  $\sim$30~keV, thus excluding the presence of the third {\sc gabs} component required in the fit with the 
{\sc highecut*pow}. We thus conclude that the {\sc gabs} component at $\sim$30~keV is likely model-dependent, and should be investigated 
further with deeper \nustar\ observations of the source. 
Depending on the specific model considered, we measured a ratio of the normalization of the Fe K$\beta$ to Fe K$\alpha$ lines 
ranging from 0.20 to 0.23, in agreement with the expectations for neutral iron \citep{palmeri03}. The measured equivalent widths of the 
Fe K$\alpha$ feature is also in line with the expected values for highly absorbed high mass X-ray binaries and other SyXBs \citep{gimenez15}.

We also attempted a pulse phase-resolved spectral analysis of the \nustar\ data to 
search for possible changes in at least the main absorption feature at $\sim$21~keV. We extracted the spectra during the phase intervals 
0.5-0.8 and 0.0-0.2 in Fig.~\ref{fig:nustar_pulse}, i.e., corresponding to the most prominent and the secondary peak of the source pulse profile. 
We could not find evidence for significant spectral changes, although the statistics of the \nustar\ spectrum extracted during the secondary peak 
was far too low to perform a detailed investigation.  

\begin{table}[ht!]
\centering  \scriptsize
 \caption{\igr\ broadband spectral analysis results. Model 1 is {\sc const*tbabs*tbpcf*gabs*gabs*gabs*(gauss+gauss+gauss+ highecut*pow)}, whereas model 2 is  
 {\sc const*tbabs*gabs*gabs*gabs*(diskbb+gauss+gauss+gauss+ highecut*pow)}. Model 3 corresponds to the case 
 {\sc const*tbabs*tbpcf*gabs*gabs*(gauss+gauss+gauss+ bbodyrad+Comptt)}. The parameter $N_{\rm H1}$ is the Galactic absorption column density and  
 $N_{\rm H2}$ the local absorption. The parameter $f$ is the partial covering fraction. Parameters kT$_{\rm BB}$ and R$_{\rm BB}$ are the temperature and radius of the 
 {\sc diskbb}/{\sc bbodyrad} component (the latter is given assuming a distance of 2.7~kpc and in the {\sc diskbb} case also considering 
 a face-on disk). The parameter $\sigma$ is the width of the cyclotron absorption features and $\tau$ their strength. 
 The centroid energy and the equivalent width (Eqw) of all Gaussian emission line are reported.\ The parameter $\Gamma$ is the photon index 
 and E$_{\rm cut}$, E$_{\rm fold}$ the parameters of the high energy cutoff. Parameters kT$_{\rm seed}$, kT, and $\tau_{\rm p}$ are the Comptt component. 
 Fluxes are given in the 0.5-10~keV (measured from the \xmm\ spectra) and in the 
 3-79~keV energy band (measured from the \nustar\ data) with units of 10$^{-11}$~\ferg. 
 Values in parentheses are those determined from the \nustar\ data and untied during the simultaneous fit with the \xmm\ data.}       
 \label{tab:spectra} 
 \begin{tabular}{@{}llll@{}} 
 \hline 
 \hline 
 \noalign{\smallskip} 
 Parameter   & Model 1  & Model 2  & Model 3 \\ 
  \noalign{\smallskip} 
  \hline
  \noalign{\smallskip} 
  $N_{\rm H1}$ (10$^{22}$~cm$^{-2}$) & 0.3 (fixed) & 48.9$^{+0.3}_{-0.4}$ &  0.3 (fixed)\\
                                     &             & (38.8$^{+1.1}_{-0.7}$)   \\
  $N_{\rm H2}$ (10$^{22}$~cm$^{-2}$) & 47.4$\pm$0.3 & ---  & 48.4$\pm$0.5 \\ 
  $f$ & 0.9965$\pm$0.0005 & --- & 0.9946$\pm$0.0006 \\
      & (0.892$\pm$0.005) & --- & (0.902$^{+0.005}_{-0.009}$) \\  
  kT$_{\rm BB}$ (keV) & --- & 0.28$^{+0.01}_{-0.02}$ & 1.41$^{+0.07}_{-0.1}$ \\
  R$_{\rm BB}$ (km) & --- & 62$^{+18}_{-5}$ & 0.35$\pm$0.05 \\
                    &     & (133$^{+41}_{-6}$) & (0.38$\pm$0.03) \\
  $E_{\rm cycl1}$ (keV) & 20.9$\pm$0.2 & 21.1$\pm$0.2 & 21.6$\pm$0.2\\
  $\sigma_{\rm cycl1}$ (keV) & 4.0 (fixed) & 4.0 (fixed) & 4.6$\pm$0.2\\ 
  $\tau_{\rm cycl1}$ & 3.6$\pm$0.2 & 3.7$\pm$0.7 & 4.4$^{+3.4}_{-0.2}$ \\  
  $E_{\rm cycl2}$ (keV) & 31.6$^{+0.6}_{-0.7}$ & 32.3$\pm$0.9 & 37.7$^{+1.8}_{-7.5}$ \\ 
  $\sigma_{\rm cycl2}$ (keV) & 7.0 (fixed) & 7.0 (fixed) & 7.0 (fixed)\\  
  $\tau_{\rm cycl2}$ & 9.1$\pm$0.5 & 9.0$\pm$0.5 & 3.8$^{+1.5}_{-0.8}$ \\ 
  $E_{\rm cycl3}$ (keV) & 44.0$\pm$1.2 &44.4$\pm$1.2 & --- \\ 
  $\sigma_{\rm cycl3}$ (keV) & 4.0 (fixed) & 4.0 (fixed) & ---\\
  $\tau_{\rm cycl3}$ & 4.3$^{+1.0}_{-0.9}$ & 4.2$^{+1.0}_{-0.9}$ & --- \\
  $E_{\rm Fe K\alpha}$ (keV) & 6.444$\pm$0.005 & 6.450$\pm$0.002 & 6.450$^{+0.003}_{-0.010}$\\
  $Eqw_{\rm Fe K\alpha}$ (keV)  & 0.19$\pm$0.01 & 0.19$\pm$0.01 & 0.19$\pm$0.05 \\
                                & (0.15$\pm$0.03) & (0.15$\pm$0.02) &  (0.14$\pm$0.04) \\    
  $E_{\rm Fe K\beta}$ (keV)  & 7.104$\pm$0.008 & 7.104$\pm$0.008 & 7.099$\pm$0.005 \\ 
  $Eqw_{\rm Fe K\beta}$ (keV)  & 0.049$^{+0.005}_{-0.017}$ & 0.049$^{+0.003}_{-0.016}$ & 0.052$^{+0.005}_{-0.012}$\\  
  $E_{\rm S K\alpha}$ (keV)  & 2.33$\pm$0.03 & 2.33$^{+0.01}_{-0.02}$ & 2.36$^{+0.03}_{-0.02}$\\ 
  $Eqw_{\rm S K\alpha}$ (keV)  & 0.51$^{+0.06}_{-0.27}$ & 2.1$^{+13.9}_{-0.5}$ & 0.32$^{+0.29}_{-0.08}$\\   
  E$_{\rm cut}$ (keV) & 5.8$\pm$0.1 & 5.7$\pm$0.1 & --- \\
  E$_{\rm fold}$ (keV) & 14.5$\pm$0.3 & 14.9$\pm$0.2 & --- \\  
  $\Gamma$ & 0.58$^{+0.03}_{-0.06}$ & 0.60$^{+0.03}_{-0.01}$ & --- \\
  kT (keV) & --- & --- &  120.6$^{+4.7}_{-1.5}$  \\
  kT$_{\rm seed}$ (keV) & --- & --- &  3.2$^{+0.4}_{-0.2}$ \\
  $\tau_{\rm p}$ & --- & --- &  0.030$^{+0.10}_{-0.003}$  \\  
  $F_{\rm 0.5-10 keV}$ & 6.38$^{+0.02}_{-0.10}$ & 6.38$^{+0.01}_{-0.11}$ & 6.38$^{+0.10}_{-0.25}$\\ 
  $F_{\rm 3-79 keV}$ & 34.5$^{+0.3}_{-1.0}$ & 34.5$^{+0.5}_{-0.9}$ & 35.3$^{+1.0}_{-1.9}$ \\    
  $\chi^{2}_{\rm red}$/d.o.f & 1.15/1031 & 1.17/1029 & 1.16/1030 \\ 
  \noalign{\smallskip}
  \hline
  \end{tabular} 
  \end{table}

\section{Optical data}
\label{sec:others} 

We obtained two spectra of the source with the Goodman Spectrograph \citep{clemens04} on the SOAR telescope, using a 0.95$''$ slit in both cases. 
The first spectrum was acquired on 2017 August 25.099 for a total exposure time of 900~s and used the 400 l mm$^{-1}$ grating, 
covering the wavelength range 4800--8820~\AA\ and providing a resolution of 5.6~\AA. The second spectrum was obtained on  
2017 August 31.097 (total exposure time 1200~s), using a higher resolution 2100 l mm$^{-1}$ grating, which provides a wavelength coverage between 6140--6700~\AA\ 
and a resolution of 0.75~\AA. Both spectra were analyzed with standard techniques. The 400 l mm$^{-1}$ spectrum was calibrated in flux using a 
first order correction for slit losses and scaled using the $i'$ magnitude at a similar epoch of that reported by 
\citet{russell17} using photometric measurements.

\subsection{Spectral analysis}
\label{sec:optical_spe} 

\begin{figure*}
  \includegraphics[width=1.0\textwidth]{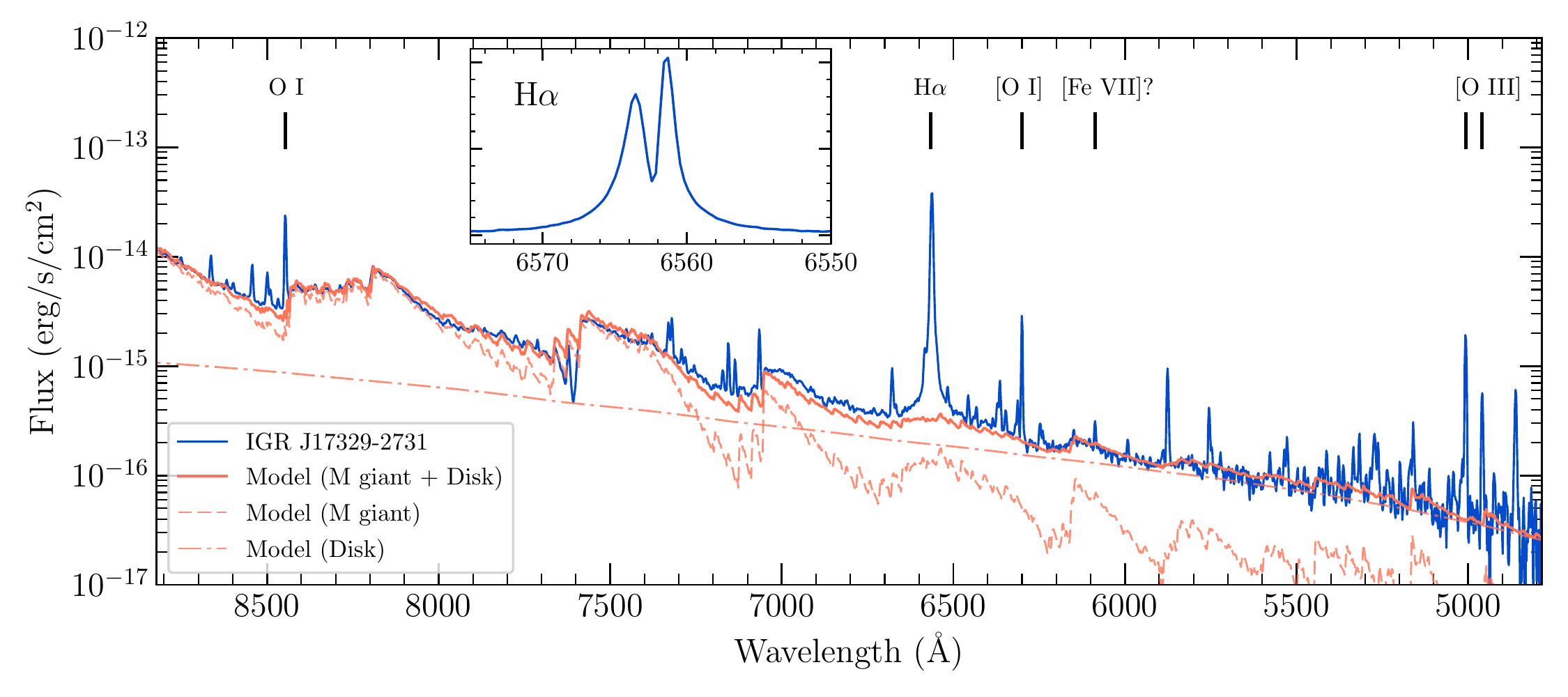}
  \caption{SOAR Goodman spectrum of \igr,\ together with a model comprising an M giant star based on Phoenix models \citep{husser13} plus an 
  accretion disk. The model spectrum was reddened assuming E(B-V)$\sim$3. The inset shows the double-peaked H$\alpha$ as seen in the high 
  resolution spectrum, indicating the presence of an accretion disk around the compact object in \src.\ }      
  \label{fig:soarspec}
\end{figure*}

The broadband optical spectrum of \igr{} contains multiple emission lines, of which the most evident is the H$\alpha$ line (Fig.~\ref{fig:soarspec}). 
The high resolution spectrum shows that this line is clearly double peaked (Fig.~\ref{fig:soarspec}). 
To estimate the properties of the observed H$\alpha$ feature in our spectrum, we deblended 
the peaks by fitting two profiles on top of a continuum. We note that the extended wings of the features are better described by Lorentzian 
profiles rather than Gaussian profiles. We estimate a peak separation of 113$\pm$2~km~s$^{-1}$. The wings of the feature are significantly broad extending 
at least $\sim$2500~km~s$^{-1}$ from the rest wavelength and are difficult to estimate accurately. The double-peaked H$\alpha$ line 
strongly suggests the presence of an accretion disk around the compact object.

The broadband spectrum shows strong TiO bands, typically observed from late M stars. The sharp TiO band 
at $\sim$8200~\AA\ allows us to classify the star as a M giant and rule out the M dwarf case.
The spectrum is fully consistent with that expected from an M giant even above 7500~\AA, but it shows a brighter continuum than expected 
for an M giant in the bluer part with only marginal evidences of TiO bands. This could be related to the contribution from the accretion disk. 
In order to test this hypothesis and simultaneously constrain the temperature of the companion, we created a spectral model comprising a synthetic 
stellar component based on the Phoenix models \citep{husser13} and an accretion disk. The effect of reddening on the complete model was also included. 
This model provided a good match with the observations, proving that the accretion disk is the dominant component at $<$7500~\AA\ 
and suggesting a temperature of the companion of $\sim$3000~K (Fig.~\ref{fig:soarspec}). This is compatible with a late M giant star \citep{star}.   
We also estimated E(B-V)$\sim$3. This value is much lower than that derived from the estimated extinction in X-rays ($\sim$46), even considering 
the most recently determined empirical correlations between E(B-V) and $N_{\rm H}$ \citep{Foight16, Bahramian15}. As it is commonly the case in strongly obscured 
X-ray binaries, this confirms that the absorption is local to the source \citep[see, e.g.,][]{walter03,chaty04,chaty12}. 

The SOAR spectra also allowed us to constrain the radial velocity of the identified oxygen and iron emission lines, leading to an estimate of 
\src\ heliocentric radial velocity of $116\pm6$ km $s^{-1}$.

\subsection{Distance to the source}

The identification of the optical counterpart to \igr\ in the Faulkes and SOAR observations also allowed us to confirm the 
association of the infrared counterpart from the 2MASS catalog 2MASS\,17325067-2730015 with magnitudes 
J=9.823, H=8.436, and K=7.775. As \igr\ was never detected in X-rays before August 2017, we assume that the 2MASS magnitudes 
were measured during quiescence and thus are dominated by the emission from the companion star. The absolute magnitudes of a late M giant 
are not well constrained, but available estimates suggest $M_K\sim -5.5$ \citep{niko00}. 
From the E(B-V) value estimated above, we thus obtain $A_K\sim1.14$ \citep[e.g., see][]{call04} and a distance to the source of  
$\sim$2.7~kpc. This value has to be taken with caution, given the limited knowledge on the absolute magnitudes of M giant stars 
and the relatively large uncertainties affecting the reddening of \igr.\ Assuming a conservative range for E(B-V), spanning from 1.0 to 4.0,  
and for M$_K$, spanning from -4.5 to -6.5, we conclude that the uncertainty associated with our estimate of the distance to the source is 
2.7$^{+3.4}_{-1.2}$~kpc.

\section{Discussion}
\label{sec:discussion}

In this paper, we reported on the data obtained from the multiwavelength campaign triggered after the discovery of the 
\inte\ transient \src.\ The analysis of the SOAR/GOODMAN data acquired by pointing at the optical counterpart of the 
source reported by \citet{russell17} led to the identification of a M6 III star at a distance of 2.7$^{+3.4}_{-1.2}$~kpc, 
thus classifying \src\ as a new member of the so-called symbiotic X-ray binaries (SyXBs). 

These systems are rare low mass X-ray binaries (LMXBs) and in some of these systems a highly magnetized ($\gtrsim$10$^{12}$~G) NS 
is accreting from the slow wind of its giant companion. These systems are characterized by the longest orbital periods among the known LMXBs (several tens to thousand 
days) and remarkably slow pulse periods that range from hundred of seconds to hours. About ten objects were discovered in this 
class so far \citep{masetti06, masetti07, masetti07b, nucita07, corbet08, nespoli08, marcu11, bozzo13, enoto14, arash14}. Detailed 
evolutionary studies on the formation channels leading to syXBs were presented by \citet{postnov11, lu12, kuranov15}. 
These studies were particularly focused on the necessary conditions to develop the long spin periods. 

The general assumption is that the progenitor of a SyXB is a binary system initially hosting a NS, endowed with a moderately strong magnetic field 
(a few $\times$10$^{12}$~G) and a spin period $\sim$0.01~s, and a low mass 
main sequence star ($\lesssim$2~M$_{\odot}$) in a wide orbit (hundred of days). In the early stages of the evolution, the 
secondary star is in the first giant branch but does not fill its Roche lobe, leading to a negligible mass transfer toward the 
NS and thus to negligible high energy emission. When this is no longer the case, the interaction of the  
NS with the surrounding material produces substantial changes in the observable properties of the system. As the magnetic field of the 
NS has not decayed and its rotational velocity is still high owing to the low spin period, the system enters in a so-called propeller phase where 
material is ejected from the vicinity of the NS by the centrifugal force at the magnetospheric boundary. This occurs at the expense of the 
rotational energy, inducing a rapid increase of the NS spin period. Depending on a number of still poorly known parameters, the rate at which the 
NS spins down and the resulting high energy emission due to the friction with the material lost by the secondary star can vary by a factor of several 
\citep{davies81, bozzo08, shakura12}. But it is relatively well understood that under these circumstances the NS can reach spin periods 
as long as $\gtrsim$10000~s. As little to no accretion is taking place in this stage, the system is still hardly detectable in the X-ray domain. 
The situation changes when the NS is slowed down sufficiently to reduce the centrifugal force at the magnetospheric boundary and allow some 
accretion to take place, finally shining as a SyXB. In this phase, accretion is still likely to take place directly from the stellar wind. 
As this is endowed with little to no angular momentum, the spin period of the NS is still expected to increase because of the effect of the friction 
between the relatively large magnetosphere and the surrounding dense environment. 
While the M giant climbs the first giant branch, the velocity of its stellar wind decreases to a level at which the formation of an accretion disk around the 
compact object is inevitable \citep{wang81}. Accretion in this phase leads to rapid decrease of the NS spin period and the system ceases to resemble 
a SyXB. At this stage, the system is few 10$^9$~yr old and moves toward a configuration that resembles more a common LMXB with a fast spinning compact object 
\citep[see Fig.~1 of][]{lu12}.  

The classification of \src\ as a SyXB with a NS accretor is strongly supported by the results of the X-ray data analysis. The prominent 
periodic modulation detected in both the \xmm\ and \nustar\ data with a period of $\sim$6680~s can be interpreted as the  
NS spin, which is in the expected range for a SyXB \citep[see, e.g., the cases of 4U\,1954+31 and IGR\,J16358-4726;][]{corbet08, patel04}. 
The energy resolved pulse profiles of \src\ are strongly reminiscent of those observed in 
other SyXB \citep[see, e.g., the case of 4U\,1954+31, showing structured pulses and a pulse fraction as high as 60-80\%;][]{enoto14} 
and high mass X-ray binaries hosting highly magnetized compact objects. In all these systems, the NS magnetic field dominates 
the dynamic of the accretion flow close to the compact star surface and the X-ray emission produced by the accretion is reprocessed within the accretion 
column(s) before reaching the observer \citep[see, e.g.,][for a recent review]{caballero12}. The structured pulse profiles can thus give an indication of the 
topology of the NS magnetic field, even though the inversion between the pulse profile and the configuration of the magnetic field is often very complex 
\citep[see, e.g.,][]{lines}. The high pulse fraction measured from the \xmm\ data ($\gtrsim$70\%) would also be compatible 
with that expected in case of a highly magnetized NS. 

Additional indications supporting the presence of a NS accretor in \src\ are obtained from the spectral analysis of the \xmm,\ \nustar,\ and 
\inte\ data. The fits to these data revealed that the relatively hard continuum, extending up to $\sim$80~keV, could be well fit with a model 
featuring a power law and a high energy cutoff, and have values similar to those used for other SyXBs \citep[see, e.g., the 
broadband spectral fits to the X-ray emission of 4U\,1954+31 reported in][]{masetti07b, enoto14} and binary systems hosting strongly magnetized 
NS \citep[see, e.g.,][and references therein]{walter2015}. The prominent and thin neutral iron K$\alpha$ and K$\beta$ lines are typical of wind-fed systems and 
are produced by the fluorescence of the X-rays due to the accretion onto the stellar wind material surrounding the compact object. The bulk of this material 
is usually spread around the entire binary, and thus often not sufficiently heated up to lead to the detection of emission lines due to more ionized 
iron states. We note that the hint in Fig.~\ref{fig:fract} of a lower pulsed fraction at energies comparable to that of the fluorescence lines further supports 
this conclusion. The source spectrum was also characterized by the presence of an evident line at 2.3~keV, which was interpreted as being the 
S K$\alpha$ line and it was already observed in a number of other wind-fed binaries with NS accretors \citep[see, e.g.,][and references therein]{fuerst11}. 
The soft excess at energies $\lesssim$2~keV could be equally fit with either a partial covering model or a disk blackbody component. The partial 
covering model is commonly used in wind-fed system and SyXBs \citep[see, e.g.,][]{masetti07b} to describe a portion of the X-ray radiation that escapes 
the local absorption due to the inhomogeneous stellar wind environment and is affected only by the Galactic extinction in the direction of 
the source \citep[see, e.g.,][]{nunez17}. The parameter $f$ 
in this model represents the fraction of the total radiation that escapes the local absorption, and the fact that we measured a significant change in $f$ 
between the non-simultaneous \xmm\ and \nustar\ data (see Table~\ref{tab:spectra}) is compatible with the idea that the accretion environment around 
the NS changes on short timescales and is mostly regulated by the physical conditions of the stellar wind. A similar conclusion applies in case the 
alternative model with a blackbody component coming from the hot spot of the NS and a Comptonized component 
from the NS accretion column(s) is considered. This model is widely used in case of strongly magnetized pulsars \citep[see, e.g.,][and references 
therein]{walter2015}, and also requires the presence of a partial covering with a variable parameter $f$ between the \xmm\ and \nustar\ data 
(although in this case the changes are less prominent).  

Another important property of the source revealed by the X-ray spectral analysis is the presence of cyclotron lines. The absorption feature around $\sim$21~keV 
is far the most prominent in the spectrum, and our fits revealed that it is unlikely to originate from a non-optimal fit of the continuum. If this absorption 
feature is interpreted as the fundamental cyclotron harmonic, then we can use the centroid energy obtained from the fit to estimate the NS magnetic field $B$ with the 
equation $B$/(10$^{12}$~G)=$E_{\rm cycl}$/11.6~keV$\times$(1+z), where $(1+z)$ is the gravitational redshift \citep[for a standard NS with a mass of 1.4~M$_{\odot}$ 
and a radius of 10~km, z$\simeq$0.3; see, e.g.,][]{caballero12}. In the case of \src,\ we can thus estimate that $B$$\simeq$2.4$\times$10$^{12}$~G is compatible with the 
value expected for a SyXB. Our spectral fit revealed two other possible absorption features at energies $\sim$30~keV and $\sim$40~keV. The statistics of the \nustar\ 
data at these energies is far from optimal, but we checked that residuals resembling two absorption features are left at these energies even when different 
choices of the background and source extraction regions are considered. We cautioned, however, that these two additional features are likely model-dependent 
and should be confirmed by future \nustar\ observations with a better S/N; the feature at $\sim$40~keV was only marginally required in the alternative 
fit with the hot {\sc bbodyrad} and the {\sc comptt} component and no significant residuals were found in this fit around $\sim$30~keV. 
If we interpret these features as additional cyclotron lines, we might be observing the first harmonic of the 
fundamental line ($\sim$40~keV) and another cyclotron feature that is not in harmonic ratio with the others ($\sim$30~keV). Although this is not always observed 
in accreting X-ray pulsars, there is at least one other example where cyclotron features at independent energies were found and ascribed to the presence 
of emission arising in distinct regions \citep[e.g., two regions at different heights from the compact object surface in the same accretion column 
or two accretion columns;][]{iyer15}. Future \nustar\ observations endowed with a higher S/N will be helpful to confirm the detection of all absorption 
features and study in more detail the possibly complex topology of the magnetic field in \src.\ Although the feature at 30~keV remains to be investigated, 
at the best of our knowledge, \src\ is the first SyXB for which a direct measurement of the NS magnetic field is reported. 

The presence of a disk around the compact object in \src\ is suggested by the optical spectrum (see Sect.~\ref{sec:optical_spe}), 
but we could not find  strong evidence for the presence of this disk in the source X-ray spectrum. We included for completeness in Table~\ref{tab:spectra} 
the fit to the source X-ray spectra including a {\sc diskbb} component in place of the partial covering, but the latter model remained slightly more 
statistically preferable. 
Assuming the most probable distance to the source of 2.7~kpc and using the 3-79~keV X-ray flux reported in Table~\ref{tab:spectra}, we can estimate that at the 
time of the \xmm\ and \nustar\ observations the luminosity of the source was $L_{\rm X}$=3$\times$10$^{35}$~erg~s$^{-1}$ and thus the mass accretion rate onto the NS was 
$\dot{M}$$\simeq$$L_{\rm X}$$R_{\rm NS}$/(G$M_{\rm NS}$)$\sim$1.6$\times$10$^{15}$~g~s$^{-1}$. The measurement of the NS magnetic field provided by the cyclotron line 
discussed above would imply an accretion disk truncated around $R_{\rm M}$$\sim$1.6$\times$10$^9$~cm \citep[see, e.g.,][and references therein]{bozzovietri}. This is   
hardly compatible with the disk blackbody radius reported in Table~\ref{tab:spectra}, unless the system is assumed to 
be observed almost edge-on. We did not observe eclipses from the system, but if its orbital period is as long as those usually measured in SyXBs, 
this possibility cannot be ruled out and might deserve further attention during future observational campaigns. As summarized before, the symbiotic binaries 
are known to develop disks at some point while climbing up the first giant branch owing to the decrease in the stellar wind velocity. 

A peculiar characteristic of \src\ is that, at odds with respect to the other SyXBs, the source was never detected in the past. The    
\inte\ instrument observed the region around \src\ for more than 31~Ms in the past 15~yrs, and we obtained a stringent upper limit on any previous X-ray emission 
that could have been detected from the source (see Sect.~\ref{sec:integral}); the \rosat\ data did not provide constraining upper limits (see Sect.~\ref{sec:rosat}). 
The \swift/XRT monitoring carried out for about three months after the discovery of the source 
did not show any evidence for a decline in the X-ray flux nor did it reveal significant spectral changes. We thus conclude that the most 
likely possibility is that \inte\ caught the first detectable X-ray emission from the source. According to the evolutionary scenario described earlier 
in this section, it is possible that the donor star in this system just reached the stage in which the mass transfer toward the NS is sufficiently high to trigger 
substantial accretion. The mass loss rate could be boosted due to the thermal pulses that are taking place in the later stages of a red giant 
evolution \citep[see, e.g.,][and references therein]{pols01}. These events can significantly enhance the luminosity of the star, 
providing a possible explanation for the brightening of \src\ in the R band by more than four magnitudes since 1991 (a factor of $\sim 40$ in luminosity) and 
suggesting that a similar event might have also led to the comparable brightening in 1976 (see Sect.~\ref{sec:intro}). Additional observations are being planned 
to probe this possibility more concretely.   \\
  
\noindent {\em Note added in proof}: While this paper was undergoing the refereeing process, the source \src\ was again observed with \swift/XRT starting 
from 2018 February 1. These observations carried out since this date (1~ks every two days) are part of an ongoing monitoring campaign that will last (at least) until mid-2018. 
At the moment of writing, \src\ is currently being observed by XRT at a similar flux compared to that measured shortly after the discovery. The X-ray spectral 
properties of the source did not show any major change in the XRT energy band (0.5-10~keV).

\section*{Acknowledgements}
The multiwavelength campaign promptly triggered after the discovery of \src\ was made possible 
thanks to the SMARTNet community \citep{matt17} and the available online tool (http://www.isdc.unige.ch/smartnet). 
We are grateful to the \xmm,\ \nustar,\ and \swift\ teams for the effort in scheduling the ToO  
observations of \src.\ This work made use of data 
from the {\em NuSTAR} mission, a project led by  the California Institute of Technology, managed 
by the Jet Propulsion  Laboratory, and funded by the National Aeronautics and Space Administration.  
This research has made use of the {\em NuSTAR} Data Analysis Software (NuSTARDAS) jointly 
developed by the ASI  Science Data Center (ASDC, Italy) and the California Institute of  Technology 
(USA). We also made use of observations obtained with \xmm\ and \inte,\ which are two ESA science missions with instruments and 
contributions directly funded by ESA Member States and NASA. 
The paper is also based on observations obtained at the Southern Astrophysical Research (SOAR) telescope, which is a joint project of the Minist\'{e}rio 
da Ci\^{e}ncia, Tecnologia, Inova\c{c}\~{a}os e Comunica\c{c}\~{a}oes (MCTIC) do Brasil, the U.S. National Optical Astronomy 
Observatory (NOAO), the University of North Carolina at Chapel Hill (UNC), and Michigan State University (MSU).
The Faulkes Telescope Project is an education partner of LCO. The Faulkes Telescopes are maintained and operated by LCO. 
EB and PR acknowledge financial contribution from the agreement ASI-INAF I/037/12/0. 
We acknowledge financial contribution from the agreement ASI-INAF I/037/12/0. 
AB thanks T.J.Maccarone and R. Wijnands for helpful discussions. 
AP acknowledges funding from the EUs Horizon 2020 Framework Programme for Research and Innovation under
the Marie Skłodowska-Curie Individual Fellowship grant agreement 660657-TMSP-H2020-MSCA-IF-2014. 
JS acknowledges support from the Packard Foundation.
We thank the anonymous referee for detailed comments that helped us improve the paper.

\bibliography{reference.bib}{}
\bibliographystyle{aa}

\end{document}